\documentclass[11pt]{article}

\usepackage{epsfig}
\usepackage{times}
\usepackage{epstopdf}
\usepackage{amsmath}
\usepackage{amssymb}
\usepackage[left=1.75cm,right=1.75cm,top=3.0cm,bottom=3.0cm]{geometry}
\usepackage{setspace}

\begin{document}

\bigskip\bigskip
\centerline {\Large \bf {On the Existence and Utility of Rigid Quasilocal Frames}}

\bigskip\bigskip
\centerline{\large Richard Epp$^{a}$, Robert B. Mann$^{a,b}$ and Paul McGrath$^{a}$}

\bigskip\bigskip
\centerline{${}^a$ \em Department of Physics and Astronomy, University of Waterloo, Waterloo, Ontario N2L 3G1, Canada}

\vspace{0.2 cm}
\centerline{${}^b$ \em Perimeter Institute for Theoretical Physics, Waterloo, Ontario N2L 2Y5, Canada}

\vspace{0.3cm}
\centerline{\em rjepp@uwaterloo.ca, rbmann@uwaterloo.ca, pmcgrath@uwaterloo.ca}

\bigskip\bigskip

\begin{abstract}
The notion of a \textit{rigid quasilocal frame} (RQF) provides a geometrically natural way to define a system in general relativity, and a new way to analyze the problem of motion. An RQF is defined as a two-parameter family of timelike worldlines comprising the boundary (topologically $\mathbb{R}\times\mathbb{S}^{2}$) of the history of a finite spatial volume, with the rigidity conditions that the congruence of worldlines be expansion- and shear-free.  In other words, the \textit{size} and \textit{shape} of the system do not change.

In previous work, such quasilocally-defined systems in Minkowski space were shown to admit precisely the same six degrees of freedom of rigid body motion that we are familiar with in Newtonian space-time, without any constraints, circumventing a century-old theorem due to Herglotz and Noether.  This surprising result is a consequence of the fact that any two-surface with $\mathbb{S}^{2}$ topology always admits precisely six conformal Killing vector fields, which generate an action of the Lorentz group on the sphere.  Several representative examples of RQFs were constructed, including one that provides a quasilocal resolution to the Ehrenfest paradox, also a century old.

Here we review the previous work in flat spacetime and extend it in three directions: (1) Using a Fermi normal coordinates approach, we explicitly construct, to the first few orders in powers of areal radius, the general solution to the RQF rigidity equations in a generic curved spacetime, and show that the resulting RQFs possess exactly the same six motional degrees of freedom as in flat spacetime; (2) We discuss how RQFs provide a natural context in which to understand the flow of energy, momentum and angular momentum into and out of a system; in particular, we derive a simple, exact expression for the flux of gravitational energy (a gravitational analogue of the Poynting vector) across the boundary of an RQF in terms of operationally-defined geometrical quantities on the boundary; (3) We use this new gravitational (or ``geometrical'') energy flux to resolve another apparent paradox, this one involving electromagnetism in flat spacetime, which we discovered in the course of this work.\footnote{An earlier version of this paper appeared in reference~\cite{EMM2011}. The present paper contains minor corrections and clarifications. Note: In reference~\cite{EMM2013} we changed our sign convention for the quasilocal stress to what we now believe is a more appropriate convention, but because this is not crucial in our present paper, here we continue to use our older sign convention.}

\end{abstract}

\section{Introduction \& Summary}

In 1910, Herglotz and Noether showed that a three-parameter family of timelike worldlines in Minkowski space satisfying Born's 1909 rigidity conditions does not have the six degrees of freedom of rigid body motion we are familiar with from Newtonian mechanics, but a smaller number---essentially only three~\cite{Born1909, Herglotz1910, Noether1910}. This result curtailed, to a large extent, subsequent study of rigid motion in special and (later) general relativity.

Recently we demonstrated~\cite{Epp2009} that one {\it can} implement Born's notion of rigid motion in flat spacetime---with precisely the desired three translational and three rotational degrees of freedom (with arbitrary time dependence)---provided the system is defined {\it quasilocally} as the two-dimensional set of points comprising the {\it boundary} of a finite spatial volume, rather than the three-dimensional set of points within the volume.  To accomplish this we introduced the notion of a {\it rigid quasilocal frame} (RQF) as a geometrically natural way to define a system in the context of the dynamical spacetime of general relativity.  An RQF is defined as a two-parameter family of timelike worldlines comprising the worldtube boundary (topologically $\mathbb{R}\times \mathbb{S}^{2}$) of the history of a finite spatial volume, with the rigidity conditions that the congruence of worldlines is expansion-free (the ``size'' of the system is not changing) and shear-free (the ``shape'' of the system is not changing).  In other words, the radar ranging distance between each nearest-neighbour pair of observers on the $\mathbb{S}^{2}$ boundary remains constant in time.

This quasilocal definition of a system is anticipated to yield simple, exact geometrical insights into the problem of motion in general relativity.  It begins by answering, in a precise way, the questions {\it what} is in motion (a rigid two-dimensional system boundary with topology $\mathbb{S}^{2}$, and whatever matter or radiation it happens to contain at the moment), and what motions of this rigid boundary are possible.  It also allows for a natural identification of the energy, momentum and angular momentum fluxes across the system boundary, ``natural'' in the sense that it avoids fluxes that would result from simply a change in the size or shape of the system boundary to enclose a different amount of the ambient matter or radiation.  The motion of the system can then be analyzed ``cleanly'' in terms of these natural fluxes.

In Ref~\cite{Epp2009} we constructed several representative examples of RQFs in flat spacetime.  In addition to addressing the general question of the existence of solutions to the RQF rigidity equations, two key specific results emerged: (1) The existence of precisely six time-dependent degrees of freedom in the collective motion of the RQF observers is intimately connected with the fact that any two-surface with topology $\mathbb{S}^{2}$ always admits precisely six conformal Killing vector fields, which generate an action of the Lorentz group on the sphere.  In contrast to the usual case of the Lorentz group acting locally on a single observer (rotations and boosts of his tetrad along his worldline), here we have the Lorentz group acting {\it quasi}locally on a two-sphere's worth of observers along their worldtube.  Roughly speaking, we are free to specify the six ``Newtonian'' $\ell=1$ vector spherical harmonics of the motion, but the higher $\ell$ spherical harmonics are fixed by the RQF equations, and represent relativistic corrections required to achieve this ``$\ell=1$ vector motion'' in a manner that maintains relativistic rigidity. (2) Of the familiar Newtonian motions, it is well known that the time-{\it dependent} rigid rotations are the most problematic in both special relativity (e.g., Ehrenfest's paradox~\cite{Stachel1980}) and general relativity.  Indeed, these are essentially the motions forbidden by the Herglotz-Noether theorem.  Using the RQF approach we addressed this problem and found a {\it quasilocal} resolution to Ehrenfest's paradox: roughly speaking, we showed that concentric shells of a body {\it can} be spun up rigidly, but rigidity between observers on neighbouring shells cannot be maintained.

The main purpose of the present work is twofold: (1) To show that the notion of an RQF can be easily extended from flat spacetime to a generic curved spacetime, and (2) to show that RQFs are useful for the better understanding of various fluxes, in particular gravitational energy flux. In {\S}\ref{Definition} we review the precise definition of an RQF and outline some of its basic properties.  In {\S}\ref{Intrinsic} we review the intrinsic geometry of the three-dimensional timelike worldtube of an RQF and examine the various intrinsic geometrical observables associated with an RQF.  In {\S}\ref{Extrinsic} we analyze the extrinsic geometry of an RQF and, using Brown and York's quasilocal definition of a {\it total} (matter plus gravitational) energy-momentum tensor~\cite{Brown1993}, we derive an energy conservation equation that relates the change in energy of an RQF to the usual matter energy flux, plus a certain simple, operationally-defined ``geometrical'' energy flux across the boundary.  We argue that the latter is to be interpreted as a gravitational energy flux.  In {\S}\ref{Curved} we use a Fermi normal coordinates approach to explicitly construct, to the first few orders in powers of areal radius, the general solution to the RQF rigidity equations in a generic curved spacetime. We show that the resulting RQFs possess exactly the same six motional degrees of freedom as in flat spacetime; all that changes is that the inhomogeneous terms in the rigidity equations become more complicated as they now incorporate curvature effects.  There appears to be no obstruction to iterating the solution to the RQF equations to arbitrarily high orders in powers of areal radius, which suggests that RQFs exist in a quite general context,\footnote{An important case outside of this context is the one in which the RQF contains a compact source that cannot be usefully expanded in powers of areal radius using a Fermi normal coordinates approach.  Work in this direction is currently in progress by the present authors.} and possess precisely the same degrees of freedom of motion as rigid bodies in Newtonian space-time.

In {\S}\ref{Curved} we also analyze the energy conservation equation introduced in {\S}\ref{Extrinsic} in the context of this general solution, both to gain insight into the nature of this new geometrical energy flux, and to provide evidence for its interpretation as a gravitational energy flux.  Three key results emerge: (1) Even in flat spacetime there is a nonzero geometrical energy flux, which has the form of the cross product of the RQF's $\ell=1$ acceleration and rotation rate.  This flux integrates to zero over the closed surface of the RQF (as it should), and can be interpreted, via the equivalence principle, in terms of the gravitoelectromagnetic analogue of the Poynting vector familiar from linearized gravity.  It represents a form of gravitational energy flowing through the RQF system boundary analogous to the electromagnetic energy flowing through a sphere containing crossed, static electric and magnetic dipoles in electromagnetism.  (2) We construct a simple apparent paradox involving an accelerating box in flat spacetime, immersed in a uniform electric field, with co-moving observers attempting to understand the changing electromagnetic energy inside the box in terms of the Poynting vector flux through the boundary of the box.  The paradox is that considering only the Poynting flux accounts for only {\it half} of the rate of change of electromagnetic energy inside the box.  We show that an extra geometrical energy flux which naturally appears in the RQF approach precisely resolves this discrepancy.  We interpret this extra flux as a flux of gravitational energy due to the curvature of spacetime produced by the uniform electric field, which cannot be properly understood in the context of special relativity.  (3) Pushing the analysis to the next higher order in powers of areal radius (orders now involving terms quadratic in the curvature), we find that the geometrical energy flux takes the form of the cross product of the electric and magnetic parts of the Weyl tensor, which we take as additional evidence for the interpretation of the geometrical energy flux as a gravitational energy flux.  Finally, at the next higher order, the geometrical energy flux takes a form that bears some resemblance to (but is different from) the time derivative of the ``0000'' component of the Bel-Robinson tensor. So in general, the expansion of the geometrical energy flux in powers of areal radius results in an infinite series of increasingly complicated curvature tensor terms. We believe this explains why various curvature tensor expressions have, over the years, been tentatively associated with gravitational energy, but that ultimately the curvature tensor is {\it not} the correct language for gravitational energy. We contend that the correct language is the simple coupling between the intrinsic and extrinsic curvature of an RQF represented by the geometrical energy flux, which we believe is the true gravitational energy flux.

\section{Definition of a Rigid Quasilocal Frame}\label{Definition}

Let us begin by introducing some notation.  Consider a smooth four-dimensional manifold, $M$, endowed with a Lorentzian spacetime metric, $g_{ab}$, with signature $+2$.  Naturally associated with $g_{ab}$ is its torsion-free, metric-compatible covariant derivative operator, $\nabla_{a}$, and volume form, $\epsilon_{abcd}$.  Let $\mathcal B$ denote a two-parameter family of timelike worldlines with topology $\mathbb{R}\times \mathbb{S}^{2}$, i.e., a timelike worldtube that represents the history of a two-sphere's-worth of observers bounding a finite spatial volume.  Let $u^a$ be the future-directed unit vector field tangent to this congruence, representing the observers' four-velocity.  The spacetime metric, $g_{ab}$, induces on $\mathcal B$ a spacelike outward-directed unit normal vector field, $n^{a}$, and a Lorentzian three-metric, $\gamma_{ab}:=g_{ab}-n_{a}n_{b}$.  At each point $p\in {\mathcal B}$ we have a {\it horizontal} subspace, $H_p$, of the tangent space to $\mathcal B$ at $p$, consisting of vectors orthogonal to both $u^a$ and $n^a$.  Let $\sigma_{ab}:=\gamma_{ab}+u_a u_b$ denote the spatial two-metric induced on $H:=\bigcup_p H_p$.  Finally, let $\epsilon_{ab} := \epsilon_{abcd}u^c n^d$ denote the corresponding volume form associated with $H$.  The time development of our congruence is described by the tensor field $\theta_{ab}:=\sigma_a^{\phantom{a}c}\sigma_b^{\phantom{b}d}\nabla_c u_d$.  We adopt the usual terminology: the {\it expansion} is $\theta := \sigma^{ab}\theta_{ab}$ (the trace part); the {\it shear} is $\theta_{< ab >} := \theta_{(ab)}-\frac{1}{2}\theta\sigma_{ab}$ (the symmetric trace-free part, here and elsewhere denoted by angle-brackets); and the {\it twist} is $\nu:=\frac{1}{2}\epsilon^{ab}\theta_{ab}$ (the antisymmetric part).

A {\it rigid quasilocal frame} is defined as a congruence of the type just described, with the additional conditions that the expansion and shear both vanish, i.e., the size and shape, respectively, of the boundary of the finite spatial volume---as seen by our observers---do not change with time:
\begin{equation} \label{eq:RigidityCondition}
\theta = \theta_{<ab>} = 0
\;\;\;\Longleftrightarrow\;\;\; \theta_{(ab)}=0 .
\end{equation}
These three differential constraints ensure that $\sigma_{ab}$ is a well defined two-metric on the quotient space of the congruence, $\mathcal Q\simeq \mathbb{S}^{2}$, i.e., the space of the observers' worldlines.  It describes the intrinsic geometry of the rigid ``box'' bounding the volume, as measured locally by our two-sphere's-worth of observers.  Notice that there is no restriction on $\nu$---the twist of the congruence---since we want to allow for the possibility of our rigid box to rotate, in which case the subspaces comprising $H$ are not integrable, i.e., $u^a$ is not hypersurface orthogonal as a vector field in $\mathcal B$.

For future reference, we define the observers' four-acceleration as $a^a:=u^b\nabla_b u^a$, whose projection tangential to $\mathcal B$ is $\alpha^a:=\sigma^a_{\phantom{a}b}a^b$ (an intrinsic geometrical variable), and whose projection normal to $\mathcal B$ is $n\cdot a := n_a a^a$ (an extrinsic geometrical variable).

\section{RQF Intrinsic Geometry and Intrinsic Degrees of Freedom}\label{Intrinsic}
To further clarify the RQF construction, and to establish notation for some of the results in subsequent sections, let us introduce a coordinate system adapted to the congruence.  (We will also henceforth explicitly display the speed of light, $c$, and Newton's constant, $G$.)   Thus, let two functions $x^i$ on ${\mathcal B}$ locally label the observers, i.e., the worldlines of the congruence. Let $t$ denote a ``time'' function on ${\mathcal B}$ such that the surfaces of constant $t$ form a foliation of ${\mathcal B}$ by two-surfaces with topology $\mathbb{S}^{2}$.\footnote{These two-surfaces need not be spacelike.}  Collect these three functions together as a coordinate system, $x^\mu := (t,x^i)$, and set $u^\mu := N^{-1}\delta_t^\mu$, where $N$ is a lapse function ensuring that $u\cdot u=-c^2$. The general form of the induced metric $\gamma_{ab}$ then has adapted coordinate components:
\begin{equation} \label{eq:InducedMetric}
\gamma_{\mu\nu} = \left(
\begin{array}{cc}
-c^2N^2 & Nu_j \\
Nu_i & \sigma_{ij}-\frac{1}{c^2}u_i u_j
\end{array}
\right).
\end{equation}
Here $\sigma_{ij}$, and the shift covector $u_i$, are the $x^i$ coordinate components of $\sigma_{ab}$ and $u_{a}$, respectively.  (We remind the reader that because $u^\mu := N^{-1}\delta_t^\mu$ instead of $u_\mu := -c^2 N\delta^t_\mu$, this is {\em not} an ADM decomposition of $\gamma_{ab}$.)  Note that $\sigma_{ij}\,dx^i\,dx^j$ is the radar ranging, or orthogonal distance between infinitesimally separated pairs of observers' worldlines, and it is a simple exercise to show that the RQF rigidity conditions in equation~(\ref{eq:RigidityCondition}) are equivalent to the three conditions $\partial\sigma_{ij}/\partial t =0$.  The resulting time-independent $\sigma_{ij}$ is the metric induced on ${\mathcal Q}\simeq \mathbb{S}^{2}$.

In other words, an RQF is a {\it rigid} frame in the sense that each observer sees himself to be permanently at rest with respect to his nearest neighbours.  The idea is that this is true even if, for example, a gravitational wave is passing through the RQF, in which case neighboring observers must generally undergo different proper four-accelerations in order to maintain nearest-neighbour rigidity.  They will also, in general, observe different precession rates of inertial gyroscopes (defined in the next section).  Indeed, these inertial accelerations and precession rates encode information about both the motion of their rigid box and the nontrivial nature of the spacetime it is immersed in.

It is not obvious that the rigidity conditions in equation~(\ref{eq:RigidityCondition}) can, in general, be satisfied.  In Ref~\cite{Epp2009} we addressed the question of existence of solutions by constructing representative nontrivial examples in the simplest possible context of flat spacetime.  Here we will explore this same question perturbatively in the context of a generic curved spacetime, leaving a rigorous existence proof for future work.  However, assuming that these conditions {\it are} satisfied, we are then free to perform a time-{\it in}dependent coordinate transformation amongst the $x^i$ (a relabeling of the observers) such that $\sigma_{ij}$ takes the form $\sigma_{ij}=\Omega^2 \, \mathbb{S}_{ij}$, where $\Omega^2$ is a time-independent conformal factor encoding the size and shape of the rigid box, and $\mathbb{S}_{ij}$ is the standard metric on the unit round sphere. For example, if the observers' two-geometry is a round sphere of area $4\pi r^2$, and the observers are labeled by the standard spherical coordinates $x^i=(\theta,\phi)$, then $\mathbb{S}_{ij}=\text{diagonal}(1,\sin^2\theta)$ and $\Omega=r$.  We are also free to change the time foliation of ${\mathcal B}$ such that $N=1$, i.e., $t$ is proper time for the observers.

Thus we see that the intrinsic three-geometry of an RQF has two functional degrees of freedom that---with the choice of coordinate-fixing described above---are encoded in the two components of the shift covector field, $u_j$ (which are functions of $t$ and $x^i$), as well as the time-{\it in}dependent conformal factor, $\Omega$, encoding our choice of size and shape of the rigid box.  We may also think of the dynamical degrees of freedom, $u_j$, as being encoded in the observers' (coordinate independent) proper acceleration tangential to $\mathcal B$ ($\alpha^{a}$ defined earlier), whose nonvanishing covariant components are
\begin{equation}\label{eq:ObserversAcceleration}
\alpha_{j}=\frac{1}{N}\,\dot{u}_{j}+c^2\partial_{j}\ln N,
\end{equation}
in the adapted coordinate system.  (Here an over-dot denotes partial derivative with respect to $t$, and $\partial_{j}$ denotes partial derivative with respect to $x^{j}$.)  More precisely, in addition to $\alpha_j$ we are free to specify the twist, $\nu$, on one cross section of $\mathcal B$, where
\begin{equation}\label{eq:nu}
\nu = \frac{1}{2}\epsilon^{ij}(\partial_i u_j - \frac{1}{c^2}\alpha_i u_j ),
\end{equation}
and $\epsilon^{ij}$ are the $x^i$ coordinate components of $\epsilon^{ab}$.

\section{RQF Extrinsic Geometry and Conservation Laws}\label{Extrinsic}

In this section we will review how conservation laws are usually constructed in general relativity, and how the RQF approach improves on this construction.  To begin with, it is customary to require that the matter energy-momentum tensor, $T^{ab}_\text{mat}$, be covariantly conserved: $\nabla_a T_{\rm mat}^{ab}=0$.  As is well known,~\cite{Wald1980} this condition may be interpreted as expressing local conservation of matter energy-momentum, but does not in general lead to an integrated conservation law.  An exception occurs if a Killing vector, $\Psi^a$, is present in the spacetime.  Then in the identity
\begin{equation}\label{MatterConservationEquation}
\nabla_a ( T_{\rm mat}^{ab}\Psi_b ) = ( \nabla_a T_{\rm mat}^{ab} ) \Psi_b + T_{\rm mat}^{ab} \nabla_{(a} \Psi_{b)},
\end{equation}
the two terms on the right-hand side vanish, and $J^a_\text{mat}:=-T_{\rm mat}^{ab}\Psi_b$ is a conserved current.  In other words, if $\Sigma$ is a spacelike three-surface with unit, future-directed normal vector field $\tilde{U}^a$ (where $\tilde{U}_a \tilde{U}^a =-c^2$), the quantity
\begin{equation}\label{ConservedCharge}
Q := -\frac{1}{c}\int_\Sigma \, d \Sigma \,\, \tilde{U}_a J^a_\text{mat}=\frac{1}{c}\int_\Sigma \, d \Sigma \,\, \tilde{U}_a\Psi_b T_{\rm mat}^{ab}
\end{equation}
is conserved, i.e., is independent of the choice of $\Sigma$ for fixed spatial boundary, $\partial\Sigma$, and evolves in the usual way according to a corresponding flux crossing $\partial\Sigma$.

If, for example, we are interested in an energy conservation law, which will be the primary focus of this section, there are three obvious shortcomings to this construction: (1) A generic spacetime does not admit a timelike Killing vector (or any Killing vectors, for that matter), and so such a construction is of limited value.  (2) Even if the spacetime {\it does} admit a timelike Killing vector, $\Psi^a$, and we are interested in, say, the matter energy contained in $\Sigma$, one might expect the integrand on the right-hand side of equation~(\ref{ConservedCharge}) to be $\frac{1}{c^2}\tilde{U}_a \tilde{U}_b T_{\rm mat}^{ab}$, where we have imagined a three-parameter family of observers filling the volume, whose four-velocity is taken to be $\tilde{U}^a$ (i.e., hypersurface orthogonal).  Energy is, after all, an observer-dependent quantity, and $\frac{1}{c^2}\tilde{U}_a \tilde{U}_b T_{\rm mat}^{ab}$ is the local energy density of matter as measured by observers ``at rest'' with respect to $\Sigma$.  The problem is that, even in the simplest case where $\Psi^a$ is hypersurface orthogonal, and we choose $\Sigma$ to be such an orthogonal hypersurface, i.e., the observers are moving along integral curves of the Killing vector, with their four-velocity $\tilde{U}^a$ parallel to $\Psi^a$, $\Psi^a$ is not a unit vector in general, i.e., it differs from $\tilde{U}^a$ by a nonconstant scale factor, and equation~(\ref{ConservedCharge}) is then not the expected expression for energy. (3) In any case, the matter energy in any finite volume cannot, in general, be separated out from the total energy (matter plus gravitational), and so any construction based on $T_{\rm mat}^{ab}$ is bound to be problematic at best.  We should really be seeking a {\it total} energy conservation law.  We will see in this section how the RQF approach resolves all three of these shortcomings.

In 1993,  Brown and  York~\cite{Brown1993} took an important step towards addressing the third shortcoming when they suggested, based on a careful Hamilton-Jacobi-type analysis of general relativity, that the total energy-momentum tensor (matter plus gravitational) is {\it quasilocal} in nature: it is a tensor defined in the {\it boundary} of the history of a finite spatial volume (which in our RQF approach we are denoting as $\mathcal B$), and is simply\footnote{There is the tricky question of a ``reference subtraction'' required to remove ``vacuum'' contributions to $T_{\mathcal B}^{ab}$, and also to regulate $T_{\mathcal B}^{ab}$ for infinite volumes.  However, insofar as: (1) we are dealing here with finite volumes; (2) we are primarily interested in {\it changes} in energy (and momentum and angular momentum), and so any vacuum contributions cancel out; (3) we will give some evidence that these vacuum contributions might, in fact, have some physical significance; and (4) ignoring a possible reference subtraction leads to no apparent inconsistencies or other problems, we will take the ``unreferenced'' quasilocal energy-momentum tensor in equation~(\ref{BoundarySEM}) at face value and explore the consequences.}
\begin{equation} \label{BoundarySEM}
T_{\mathcal B}^{ab} := -\frac{1}{\kappa}\,\Pi^{ab},
\end{equation}
where $\Pi_{ab}:=\Theta_{ab} - \Theta \gamma_{ab}$ is the momentum canonically conjugate to the three-metric $\gamma_{ab}$ on $\mathcal B$, $\Theta_{ab}:=\gamma_a^{\phantom{a}c}\nabla_c n_b$ is the (symmetric) extrinsic curvature of $\mathcal B$ (and $\Theta$ its trace), and $\kappa = 8 \pi G / c^4$.  For example, in the context of our RQF observers, whose four-velocity is $u^a$, the quantity ${\mathcal E}:= \frac{1}{c^2} u^a u^b T^{\mathcal B}_{ab}$ is to be considered as the quasilocal total energy {\it surface} density (energy per unit area) measured by the observers; ``quasilocal'' in the sense that it has meaning only when integrated over a closed two-surface.  Roughly speaking, and stated more precisely following equation~(\ref{LHS}) below, integrating this quasilocal energy density over a two-sphere slice of $\mathcal B$ yields a measure of the total energy (matter plus gravitational) inside any spatial volume spanning this slice.

How can we construct conservation laws associated with such a quasilocal energy-momentum tensor? An obvious approach to try is to simply write down the quasilocal analogue of equation~(\ref{MatterConservationEquation}), wherein we replace the {\it matter} energy-momentum tensor, $T_{\rm mat}^{ab}$, defined in the four-dimensional spacetime $(M,g_{ab})$, with the {\it total} energy-momentum tensor, $T_{\mathcal B}^{ab}$, defined in the three-dimensional spacetime $({\mathcal B},\gamma_{ab})$.  It might appear that this would result in conservation laws involving charges and fluxes {\it within} $\mathcal B$, which would have to be trivial because the cross sections of $\mathcal B$ are {\it closed} surfaces (two-spheres).  But because of the quasilocal interpretation of $T_{\mathcal B}^{ab}$, and the nature of the Einstein equation and the RQF conditions, we will see that we will actually be analyzing the exchange of energy, momentum and angular momentum between the RQF system and the universe external to it, i.e., charges associated with spatial volumes spanning two-sphere slices of $\mathcal B$, and fluxes passing {\it through} $\mathcal B$.

Thus, in analogy with $\Psi^a$ in equation~(\ref{MatterConservationEquation}), let $\psi^a$ be an arbitrary vector field tangent to $\mathcal B$, and consider the identity
\begin{equation}\label{Basic_Conservation_Equation}
D_a ( T_{\mathcal B}^{ab}\psi_b ) = ( D_a T_{\mathcal B}^{ab} ) \psi_b + T_{\mathcal B}^{ab} D_{(a} \psi_{b)},
\end{equation}
where $D_a$  is the covariant derivative with respect to the metric, $\gamma_{ab}$, induced in $\mathcal B$.  Integrating this identity over a portion, $\Delta {\mathcal B}$, of $\mathcal B$, between initial (${\mathcal S}_i$) and final (${\mathcal S}_f$) two-sphere slices of $\mathcal B$, we have
\begin{equation}\label{Basic_Integrated Conservation_Equation}
\frac{1}{c}\negthickspace\negthickspace\int\limits_{{\mathcal S}_f-{\mathcal S}_i}\negthickspace\negthickspace d{\mathcal S}\,\tilde{u}_{a}\psi_b T_{\mathcal B}^{ab} = -\negthickspace\int\limits_{\Delta {\mathcal B}}d{\mathcal B}\,\left[( D_a T_{\mathcal B}^{ab} ) \psi_b + T_{\mathcal B}^{ab} D_{(a} \psi_{b)}\right],
\end{equation}
where $d{\mathcal S}$ and $d{\mathcal B}$ are the volume elements on ${\mathcal S}_{i,f}$ and ${\mathcal B}$, respectively, and $\tilde{u}^a$ is the unit, future-directed vector field normal to ${\mathcal S}_{i,f}$, analogous to $\tilde{U}^a$ in equation~(\ref{ConservedCharge}).
As will be discussed in detail below, the left-hand side of equation~(\ref{Basic_Integrated Conservation_Equation}) represents the change in some physical quantity of the RQF (energy, momentum or angular momentum) between ${\mathcal S}_{i}$ and ${\mathcal S}_{f}$.  The right-hand side represents two types of flux crossing the timelike boundary, $\Delta {\mathcal B}$, spanning ${\mathcal S}_{i}$ and ${\mathcal S}_{f}$, that account for this change: a matter flux, represented by the term $( D_a T_{\mathcal B}^{ab} ) \psi_b$, and a ``geometrical,'' or gravitational, flux, represented by the term $T_{\mathcal B}^{ab} D_{(a} \psi_{b)}$.  The choice of which physical quantity we are concerned with (energy, momentum or angular momentum) depends on our choice of $\psi^a$.  (Note that $( D_a T_{\mathcal B}^{ab} ) \psi_b$ and $T_{\mathcal B}^{ab} D_{(a} \psi_{b)}$ are outward-directed fluxes that cause a decrease in the corresponding physical quantity, which explains the presence of the negative sign in equation~(\ref{Basic_Integrated Conservation_Equation}))

To begin with, let us establish some notation.  We follow Brown and York~\cite{Brown1993} in splitting the total energy-momentum tensor into physically distinct parts:
\begin{equation}\label{SurfaceSEMcomponents}
\begin{split}
{\mathcal E} & := \frac{1}{c^2} u^a u^b T^{\mathcal B}_{ab}\\
c{\mathcal P}_a & := - \frac{1}{c} \sigma_{a}^{\phantom{a}b}u^{c}T^{\mathcal B}_{bc}\\
{\mathcal S}_{ab} & := \sigma_{a}^{\phantom{a}c}\sigma_{b}^{\phantom{b}d}T^{\mathcal B}_{cd},
\end{split}
\end{equation}
which denote, respectively, the energy surface density (energy per unit area), the momentum surface density (momentum per unit area), and the spatial stress (force per unit length) as seen by our RQF observers, whose four-velocity is $u^a$.

Next, let us choose $\psi^a=u^a /c$, which will give us an energy conservation law associated with the quasilocal energy-momentum current $J^a_{\mathcal B}:=-T_{\mathcal B}^{ab}\psi_b = {\mathcal E}u^a /c + c{\mathcal P}^a$.  In this case the integrand on the left-hand side of equation~(\ref{Basic_Integrated Conservation_Equation}) becomes $\frac{1}{c^2}\tilde{u}_{a}u_b T_{\mathcal B}^{ab}$.  When the observers are ``at rest'' with respect to ${\mathcal S}_{i,f}$ (i.e., $u^a=\tilde{u}^a$), this integrand reduces to $\frac{1}{c^2}\tilde{u}_{a}\tilde{u}_b T_{\mathcal B}^{ab}=\frac{1}{c^2}u_{a}u_b T_{\mathcal B}^{ab}={\mathcal E}$, which is the expression for (now {\it total}) energy density one might expect---recall the discussion of ``shortcomings'' in the paragraph following equation~(\ref{ConservedCharge}).  In general, however, the observers are {\it not} ``at rest'' with respect to ${\mathcal S}_{i,f}$, and $u^a$ and $\tilde{u}^a$ are related by a boost transformation: $\tilde{u}^a=\alpha (u^a -\beta^a )$.  Here $\beta^a\in H$ is a ``shift'' vector that can be interpreted as the tangential, spatial two-velocity of the RQF observers ``gliding over the sphere,'' and $\alpha$ is an inverse ``lapse'' function that corresponds to the ``$\gamma$-factor'' of the associated Lorentz transformation. In this, the general case, the $-\tilde{u}_a /c$ projection of the energy-momentum current $J^a_{\mathcal B}$ suffers a Lorentz transformation, and the left-hand side (LHS) of equation~(\ref{Basic_Integrated Conservation_Equation}) becomes
\begin{equation}\label{LHS}
\text{ LHS of equation (\ref{Basic_Integrated Conservation_Equation})} =\negthickspace\negthickspace\negthickspace\int\limits_{{\mathcal S}_f-{\mathcal S}_i}\negthickspace\negthickspace\negthickspace d{\mathcal S}\,\alpha\left({\mathcal E}+\beta_{a}{\mathcal P}^{a}\right).
\end{equation}
It is worth noting that $\beta^a$ contains the same information as the dynamical degrees of freedom of the RQF intrinsic geometry, $u_j$, discussed in the previous section.  To see this, we can choose our adapted coordinate system such that ${\mathcal S}_{i,f}$ are surfaces of constant $t$, and then clearly we require $\tilde{u}_j=0$, i.e., $\beta_j = u_j$ (and note that $\beta_t =0$ by construction).

There are at least two general arguments suggesting the plausibility of ${\mathcal E}:=\frac{1}{c^2}u_{a}u_b T_{\mathcal B}^{ab}$ as a total energy surface density, and thus equation~(\ref{LHS}) as a total system energy (matter plus gravitational).  First, note that
\begin{equation}\label{SpatialExtrinsicCurvature}
{\mathcal E}:=\frac{1}{c^2}u_{a}u_b T_{\mathcal B}^{ab}=-\frac{1}{\kappa c^2}u^a u^b (\Theta_{ab}-\Theta\gamma_{ab})=-\frac{1}{\kappa}\sigma^{ab}\Theta_{ab}=-\frac{c^4}{8\pi G}k,
\end{equation}
where $k:=\sigma^{ab}\Theta_{ab}$.  Consider the simplest case, in which $u^a$ is hypersurface orthogonal (as a vector field in $\mathcal B$), and let $\mathcal S$ be a two-sphere slice of $\mathcal B$ to which $u^a$ is everywhere orthogonal (i.e., the observers are ``at rest'' with respect to $\mathcal S$).  Then $k$ is the trace of the extrinsic curvature of $\mathcal S$ in the $n^a$ direction normal to $\mathcal B$.  In other words, $k$ measures the fractional rate at which the surface area elements of $\mathcal S$ increase as we move a unit proper distance radially outwards.  For example, if $\mathcal S$ is a round sphere of areal radius $r$ in Euclidean three-space, then $k=2/r$.  If we now put some mass-energy inside the sphere, it will ``warp'' the three-space such that, for a sphere of the same areal radius, its surface area will increase {\it less rapidly} than we expect based on Euclidean intuition.  (Think of the standard funnel-shaped embedding diagram for the warped three-space surrounding a compact object---the more mass-energy inside, the steeper the slope of the funnel, and the less rapidly the area of a spatial two-sphere slice will increase as we move a unit proper distance radially outwards.) Thus, the presence of mass-energy inside the sphere {\it decreases} $k$ (increases $-k$), and in this way $\mathcal E$ is a quasilocal measure of the mass-energy inside the sphere.  This measure naturally includes a gravitational energy contribution.  See reference~\cite{Brown1993} for a detailed discussion and examples (in particular, see the discussion surrounding their equation (6.15)).

As a second general plausibility argument for the interpretation of ${\mathcal E}:=\frac{1}{c^2}u_{a}u_b T_{\mathcal B}^{ab}$ as a total energy surface density, note that $\Theta = (\sigma^{ab}-\frac{1}{c^2}u^a u^b )\nabla_a n_b = k+\frac{1}{c^2}n\cdot a$.  Defining the pressure as one-half the trace of the spatial stress, $\mathbb{P}:= \frac{1}{2}\sigma^{ab}{\mathcal S}_{ab}$, and using equation~(\ref{SpatialExtrinsicCurvature}), we have
\begin{equation}\label{EPnDOTaRelation}
{\mathcal E} = \frac{c^2}{4\pi G} n\cdot a -2 \mathbb{P}.
\end{equation}
The first term on the right-hand side is intuitively satisfying in that the normal component of acceleration, $n\cdot a$, required for observers to hover a fixed distance from a static, compact object is clearly a measure of the mass-energy of the object.  For example, in the Newtonian limit of a two-sphere's worth of observers hovering a fixed radial distance $r$ from a point mass $M$, the normal component of their acceleration is $n\cdot a = GM/r^2$, and integrating the first term in equation~(\ref{EPnDOTaRelation}) over the two-sphere yields a contribution of $Mc^2$ to the total energy.  More generally, for any spherically symmetric mass distribution, any mass outside the radius of the RQF sphere has no effect on $n\cdot a$. In other words this method of measuring mass is {\it inherently quasilocal} in nature, not referring to anything outside of the system.  And of course it is closely related to Komar's definition of mass (see, e.g., section 11.2 of reference~\cite{Wald1980} for a discussion of the Komar mass), except that here we are not restricted to stationary (and asymptotically flat) spacetimes.  So one might wonder why this is not the end of the story---why there is a $-2 \mathbb{P}$ term added to the $n\cdot a$ term.  It has to do with the distinction between (inertial) mass and energy.  To see this, it helps to rewrite the equation as $\frac{c^2}{4\pi G} n\cdot a={\mathcal E}+2 \mathbb{P}$.  It is well known that, in relativity, pressure contributes to the inertia of a system, and it is plausible that this equation is the quasilocal, {\it total} energy (matter plus gravitational) version of this phenomenon.

Let us now turn our attention to the right-hand side of equation~(\ref{Basic_Integrated Conservation_Equation}), beginning with the first term---the matter flux term.  Using the identity $D_a \Pi^{ab}=\gamma^{bc}n^{d}G_{cd}$, where $G_{cd}$ is the Einstein tensor, and the definition of $T_{\mathcal B}^{ab}$ in equation~(\ref{BoundarySEM}), this term is:
\begin{equation} \label{eq:MatterFlux}
( D_a T_{\mathcal B}^{ab} ) \psi_b=-{\frac{1}{\kappa}}n^{a} \psi^{b} G_{ab}= -n^{a} \psi^{b} T^{\rm mat}_{ab},
\end{equation}
where for the last equality we used the Einstein equation.  Unlike in equation~(\ref{MatterConservationEquation}), where $\nabla_a T_{\rm mat}^{ab}=0$, in equation~(\ref{Basic_Conservation_Equation}) the divergence of the quasilocal total energy-momentum tensor is not zero.  If $\nabla_a T_{\rm mat}^{ab}$ was not zero it would signal the presence of ``external sources,'' i.e., matter fields not accounted for in $T_{\rm mat}^{ab}$ that are nevertheless interacting with the system represented by $T_{\rm mat}^{ab}$.  In our case, the nonvanishing of $D_a T_{\mathcal B}^{ab}$ likewise corresponds to external sources, now in the form of various matter fluxes passing {\it through} $\mathcal B$ that interact with the RQF system represented by $T_{\mathcal B}^{ab}$.  These fluxes of energy, momentum, and angular momentum (depending on our choice of $\psi^a$) are intimately connected with the {\it motion} of the RQF system, by which we mean inertial accelerations and precession rates of inertial gyroscopes.

For example, let us again take $\psi^a= u^a /c$ (so we are dealing with energy conservation), and consider an electromagnetic (em) field present at $\Delta {\mathcal B}$.  Then, setting $T^{\rm mat}_{ab}$ to $T^{\rm em}_{ab}$, we have
\begin{equation}\label{Electromagnetic Flux}
-n^{a} u^{b} T^{\rm em}_{ab} =  \frac{c}{4\pi} \epsilon_{ab}e^{a}b^{b} ,
\end{equation}
where $e^a$ and $b^a$ denote the electric and magnetic fields experienced by observers with four-velocity $u^a$ on $\Delta \mathcal B$.  As expected, this expression is the $n^a$ (outward-directed) component of the Poynting vector, i.e., if electromagnetic energy is leaving the system, the energy of the RQF will decrease---the energy change in equation~(\ref{LHS}) will be negative.  If, on the other hand, $\psi^a$ is taken to be in a spatial direction (tangential to $\mathcal B$ and orthogonal to $u^a$), equation~(\ref{eq:MatterFlux}) represents forces and torques (e.g., components of the Maxwell stress tensor in the case of electromagnetism) that cause changes in the momentum and angular momentum of the RQF.\footnote{This is, of course, not surprising.  In the same way that $\nabla_a T_{\rm mat}^{ab}=0$ is intimately connected with the equations of motion of matter, equation~(\ref{eq:MatterFlux}) is intimately connected with the equations of motion of an RQF system.}  This will be discussed in more detail elsewhere.

Finally, let us turn our attention to the most interesting, and last, term on the right-hand side of equation~(\ref{Basic_Integrated Conservation_Equation}).  To begin with, recall that in equation~(\ref{MatterConservationEquation}) we required both terms on the right-hand side to vanish in order for $J^a_\text{mat}=-T_{\rm mat}^{ab}\Psi_b$ to be a conserved current.  We have just seen that in the quasilocal analogue of this equation, equation~(\ref{Basic_Conservation_Equation}), the first term on the right-hand side need {\it not} vanish in order for $J^a_{\mathcal B}=-T_{\mathcal B}^{ab}\psi_b$ to be a conserved current.  Indeed, its nonvanishing simply represents the matter flux naturally associated with the time evolution of the corresponding charge.  Together with the fact that, in the quasilocal case, we are dealing with the {\it total} energy-momentum tensor (matter plus gravitational), it is thus reasonable to guess that the second term on the right-hand side of equation~(\ref{Basic_Conservation_Equation}) need not vanish in order for $J^a_{\mathcal B}$ to be a conserved current, and that, in fact, it will represent a gravitational contribution to the associated flux.

So while it needn't, and in general doesn't, vanish, there is a good reason---that will become evident shortly---to see how far we can go to making the second term on the right-hand side of equation~(\ref{Basic_Conservation_Equation}) vanish.  In analogy with $\Psi^a$ in equation~(\ref{MatterConservationEquation}), it would obviously vanish if $\psi^a$ was a Killing vector of the boundary three-geometry.  But the boundary three-geometry will not, in general, admit any Killing vectors---and certainly not in the most interesting case when gravitational radiation is crossing the boundary.  So we will do the next best thing and ask, ``To what extent---in the {\it general} case---can $\psi^a$ have Killing vector-{\it like} properties?''

For example, while the boundary three-geometry will not, in general, admit a timelike Killing vector---a time symmetry of the {\it full} three-geometry---can the observers' congruence of worldlines always be chosen such that it admits a time symmetry of at least the spatial two-geometry of the bounding box of the RQF?  Taking $\psi^a \propto u^a$, this is equivalent to asking if the {\it spatial projection}, $\sigma_{a}^{\phantom{a}c}\sigma_{b}^{\phantom{b}d}D_{(c}u_{d)}=0$, of the boundary Killing equation can be satisfied {\it in general}.  Indeed, these are precisely the RQF rigidity conditions in equation~(\ref{eq:RigidityCondition}).  We will argue below, and in the next section, that the failure of $u^a$ to satisfy the {\it full} Killing equation in the boundary is associated in a simple and precise way with gravitational energy passing through that boundary.  Also, it is worth noting that this spatially projected Killing equation is invariant under a rescaling (``conformal'' transformation) of $u^a$ by an arbitrary function.  This is the underlying reason explaining how the RQF approach resolves the second of the three ``shortcomings'' discussed in the paragraph following equation~(\ref{ConservedCharge}).

Similarly, the boundary three-geometry will not, in general, admit a spacelike Killing vector. However, as already emphasized, the spatial two-geometry of the bounding box of an RQF always admits precisely six {\it conformal} Killing vectors (CKVs): three boost-like and three rotation-like.  Taking $\psi^a$ to be such a boost- or rotation-like CKV results in simple, exact expressions for gravitational momentum or angular momentum, respectively, passing through the bounding box (and corresponding expressions for total---matter plus gravitational---momentum or angular momentum charges).  This will be reported on in detail elsewhere.  For now we consider only the case $\psi^a \propto u^a$.

In particular, taking $\psi^a = u^a /c$, as before, it is easy to see from the definitions in equation~(\ref{SurfaceSEMcomponents}) that, when the RQF rigidity conditions in equation~(\ref{eq:RigidityCondition}) are satisfied,
\begin{equation}\label{GravitationalFluxesRQF}
T_{\mathcal B}^{ab} D_{(a} u_{b)} = \alpha_a {\mathcal P}^a .
\end{equation}
We claim that the quantity $\alpha_a {\mathcal P}^a$ is a simple, exact expression for the outward-directed ``geometrical'' flux of gravitational energy across the boundary of an RQF.  We will provide detailed evidence for this interpretation in the next section, but to immediately see that this is plausible, observe that, starting from the definition in equation~(\ref{SurfaceSEMcomponents}),
\begin{equation}\label{InterpretationOfScriptP}
{\mathcal P}^a :=-\frac{1}{c^2}\sigma^{ab}u^c T^{\mathcal B}_{bc}=\frac{1}{\kappa c^2}\sigma^{ab}u^c (\Theta_{bc}-\Theta\gamma_{bc})=\frac{c^2}{8\pi G}\sigma^a_{\phantom{a}b}u^c\nabla_c n^b=:\frac{c^2}{8\pi G}\epsilon^a_{\phantom{a}b}\omega^b ,
\end{equation}
where $\omega^a$ is the precession rate of inertial gyroscopes as measured by the RQF observers (i.e., the $\epsilon^a_{\phantom{a}b}$ projection of the rotation rate of $n^a$ under parallel transport along $u^a$).  So ${\mathcal P}^a$ can be interpreted as either a tangential (to $\mathcal B$) momentum surface density, or a gyroscope precession rate (projected tangentially to $\mathcal B$ and rotated by 90 degrees).  Hence, our geometrical, or gravitational energy flux can be expressed in the more suggestive form:
\begin{equation}\label{ACrossOmegaFormOfGeometricalFlux}
\alpha_a{\mathcal P}^a = \frac{c^2}{8\pi G}\epsilon_{ab}a^a \omega^b ,
\end{equation}
(where we have replaced, with impunity, the tangential acceleration, $\alpha^a$, with the full four-acceleration, $a^a$).  The analogy with electromagnetic energy flux is striking---recall equation~(\ref{Electromagnetic Flux}), with the identifications $e^a\leftrightarrow a^a$ and $b^a\leftrightarrow \omega^a$.  It is well known that, in the gravitoelectromagnetic interpretation of linearized general relativity, the gravitoelectric field is associated with acceleration, and the gravitomagnetic field with rotation~\cite{Mashhoon}.  So if one had to guess an expression for gravitational energy flux, analogous to that in electromagnetism, one might well try an expression like $\epsilon_{ab}a^a \omega^b$, multiplied by $c^2/G$ to get the units right. The question would then be, ``What do we put for $a^a$ and $\omega^a$?''  The RQF approach provides a natural and precise answer: $a^a$ is the concerted two-parameter family of four-accelerations that observers must undergo in order to maintain constant radar ranging distances to all nearest neighbor observers, i.e., to ``actively'' (e.g., with rockets) compensate for the geodesic deviations they would otherwise experience in freefall in a dynamical spacetime.  And $\omega^a$ is derived uniquely from the resulting $a^a$.  It is remarkable that this simple expression for gravitational energy flux is actually {\it exact}, in the full nonlinear theory.  Moreover, it is {\it operational}: RQF observers can directly measure $a^a$ and $\omega^a$ using accelerometers and gyroscopes, and thus determine the quasilocal gravitational energy flux at their respective positions.

In summary, substituting $\psi^a = u^a /c$ into the general conservation equation~(\ref{Basic_Integrated Conservation_Equation}) gives us an energy conservation law that relates the change in {\it total} energy of the RQF system (matter plus gravitational), between initial and final time slices ${\mathcal S}_{i}$ and ${\mathcal S}_{f}$ of $\mathcal B$, to two types of energy flux, matter and gravitational, through the timelike three-surface, $\Delta {\mathcal B}$, spanning ${\mathcal S}_{i}$ and ${\mathcal S}_{f}$:
\begin{equation}\label{eq:SimpleConservationEq}
\int\limits_{{\mathcal S}_f-{\mathcal S}_i} \negthickspace\negthickspace d{\mathcal S}\,\alpha\left({\mathcal E}+\beta_{a}{\mathcal P}^{a}\right)=
\frac{1}{c}
\int\limits_{\Delta {\mathcal B}} d{\mathcal B} \, \left(  n^{a} u^{b} T^{\rm mat}_{ab} - \alpha_a {\mathcal P}^a \right).
\end{equation}
On the left-hand side of this equation, $\alpha$ and $\beta_a$ represent a Lorentz boost from the observers' four-velocity, $u^a$, to the four-velocity $\tilde{u}^a$ of observers momentarily ``at rest'' with respect to ${\mathcal S}_{i}$ and ${\mathcal S}_{f}$.  In the next section we will examine this energy conservation law in detail, highlighting the interpretation and significance of the geometrical, or gravitational energy flux term, $\alpha_a {\mathcal P}^a$.

\section{RQFs in Curved Spacetime}\label{Curved}

In reference~\cite{Epp2009} we addressed the existence and utility of RQFs in flat spacetime.  Based on the results of the previous section, we will now extend this work to a generic curved spacetime, proceeding perturbatively in powers of the areal radius of a round sphere RQF.  Let us begin by writing down the components of the metric, $g_{ab}$, in Fermi normal coordinates, $X^{a} := (cT,X^{I})$, $I=1,\,2,\,3$, in the neighborhood of a timelike worldline, $\mathcal C$ (with arbitrary acceleration) in a generic spacetime:\cite{standardFNCreference}
\begin{align}
g_{00} &= - \left(1 + \frac{1}{c^2} A_{K} X^{K}\right)^2 + \frac{1}{c^2} R^2 W_{K} W_{L} P^{KL} - \overset{\mathtt{o}}{R}_{0K0L} X^{K} X^{L} + \mathcal{O}(R^3),\label{eq:FNCmetric00}\\
g_{0J} &=  \frac{1}{c} \epsilon_{JKL}  W^{K} X^{L} - \frac{2}{3} \overset{\mathtt{o}}{R}_{0KJL} X^{K} X^{L} + \mathcal{O}(R^3),\label{eq:FNCmetric0J}\\
g_{IJ} &= \delta_{IJ} - \frac{1}{3} \overset{\mathtt{o}}{R}_{IKJL} X^{K} X^{L} + \mathcal{O}(R^3),\label{eq:FNCmetricIJ}
\end{align}
where $ R^2 := \delta_{IJ} X^I X^J$, $A_{K}(T)$ is the proper acceleration along $\mathcal C$, $W_{K}(T)$ is the proper rate of rotation of the spatial axes (triad) along $\mathcal C$, $P^{KL} := \delta^{KL} - X^{K} X^{L} / R^2$ projects vectors perpendicular to the radial direction, and $ \overset{\mathtt{o}}{R}_{abcd}(T)$ are the Fermi normal coordinate components of the Riemann curvature tensor evaluated on $\mathcal C$.  Note that $T$ is the proper time along $\mathcal C$ ($R=0$), and an overset circle indicates a quantity evaluated on $\mathcal C$ (except for $A_{K}$ and $W_{K}$, which obviously refer to $\mathcal C$).

Let us now embed, into this coordinate system, a two-parameter family of worldlines in the neighborhood of $\mathcal C$ that will represent a two-sphere's worth of observers, i.e., a fibrated timelike worldtube, $\mathcal B$, surrounding $\mathcal C$.  To do this, we introduce a second set of coordinates, $x^\alpha := (t,\,r,\,x^i )$, which are the coordinates $x^\mu := (t,\,x^i )$ on $\mathcal B$ introduced in {\S}\ref{Intrinsic}, augmented by a radial coordinate, $r$.  Then, taking $x^i =(\theta,\phi)$, we introduce the coordinate transformation:
\begin{align}
T(t,\,r,\,\theta,\phi) &= t,\label{T=t}\\
X^I(t,\,r,\,\theta,\phi) &= r r^I(\theta,\phi) + r^3 f^I(t,\theta,\phi) + \mathcal{O}(r^4),\label{X^I=...}
\end{align}
where $r^{I}(\theta,\phi) := (\sin{\theta}\cos{\phi},\,\sin{\theta}\sin{\phi},\,\cos{\theta})$ are the standard direction cosines of a radial unit vector in spherical coordinates in Euclidean three-space.  The idea is that the three arbitrary functions $f^I(t,\theta,\phi)$ (and their counterparts at higher order in $r$) allow us the full freedom to ``wiggle'' the observers' worldlines (defined by $r,\,\theta,\,\phi=$ constant) arbitrarily in the three spatial directions, and are to be chosen such that the three RQF rigidity conditions in equation~(\ref{eq:RigidityCondition}) are satisfied.  Specifically, we will demand that the observers' radar ranging two-metric, $\sigma_{ij}$, induced by the embedding, be equal to $r^2 \mathbb{S}_{ij}$, so that the observers find themselves on a round sphere of areal radius $r$.  (Recall {\S}\ref{Intrinsic} for a definition of our notation.)

There are three points worth noting: (1) The RQF conditions, which are equivalent to $\partial\sigma_{ij}/\partial t=0$ in our adapted coordinate system, are clearly invariant under a time reparametrization, and so to simplify the analysis as much as possible we have chosen surfaces of constant $t$ to coincide with surfaces of constant $T$, i.e., $T=t$ in equation~(\ref{T=t}); (2) the RQF conditions are obviously trivially satisfied at lowest order ($X^I = r r^I$ in equation~(\ref{X^I=...})), and we find that the first nontrivial order is two orders of $r$ higher, which explains the absence of an $\mathcal{O}(r^2)$ term in equation~(\ref{X^I=...}); and (3) for technical reasons it proves useful to decompose $f^I$ as follows:
\begin{align}\label{f^Idecomposition}
f^I (t,\theta,\phi) = F(t,\theta,\phi) r^I (\theta,\phi) + f^{i} (t,\theta,\phi) \mathbb{B}^{I}_{i}(\theta,\phi).
\end{align}
Here $F$ encodes a radial, or normal perturbation of the observers' worldlines, and $f^i$ encodes an angular, or tangential perturbation, together comprising three functional degrees of freedom.  We have also defined the {\it boost generators} $\mathbb{B}^{I}_{i} := \partial_{i} r^I $ (more on these below), and made use of the completeness relation $\delta^{IJ}=r^I r^J + \mathbb{S}^{ij}\mathbb{B}_i^I\mathbb{B}_j^J$, where $\mathbb{S}^{ij}$ is the matrix inverse of $\mathbb{S}_{ij}$.

With this construction, we find that the induced radar ranging two-metric seen by the RQF observers is:
\begin{align}\label{DiffEqRQF}
\sigma_{ij} =& r^2 \mathbb{S}_{ij} + r^4 \left(2 \mathbb{D}_{(i} f_{j)} + 2 F \mathbb{S}_{ij} - \frac{1}{3} \overset{\mathtt{o}}{R}_{IKJL} \mathbb{B}^I_i \mathbb{B}^J_j r^K r^L  + \frac{1}{c^2} W_I W_J \mathbb{R}^I_i \mathbb{R}^J_j\right)
+ \mathcal{O}(r^5) ,
\end{align}
where here, and in what follows, the quantities $A_K$, $W_K$ and $\overset{\mathtt{o}}{R}_{abcd}$, which in equations~(\ref{eq:FNCmetric00}) to~(\ref{eq:FNCmetricIJ}) are functions of $T$, are now functions of $t$, according to equation~(\ref{T=t}).
We have defined $f_i :=\mathbb{S}_{ij}f^j$, and $\mathbb{D}_i$ is the covariant derivative operator associated with the unit round sphere metric, $\mathbb{S}_{ij}$.  Letting $\mathbb{E}_{ij}$ denote the volume form associated with $\mathbb{S}_{ij}$, we have also defined $\mathbb{R}^I_i := - \mathbb{E}_i^{\phantom{i}j} \mathbb{B}^I_j$ as the {\it rotation generator} counterparts to $\mathbb{B}^{I}_{i}$.  The contravariant form of these generators is given by $\mathbb{B}_I^i = \delta_{IJ} \mathbb{S}^{ij} \mathbb{B}_j^J$ and $\mathbb{R}_I^i = \delta_{IJ} \mathbb{S}^{ij} \mathbb{R}_j^J$.  It is easy to verify that the six vector fields $\mathbb{B}_I^i \partial_i$ and $\mathbb{R}_I^i \partial_i$ are conformal Killing vectors on the unit round sphere in Euclidean three-space, and that their commutators yield a representation of the Lorentz algebra.

Inspection of equation~(\ref{DiffEqRQF}) reveals that to satisfy the RQF rigidity conditions ($\sigma_{ij}=r^2 \mathbb{S}_{ij}$) to lowest nontrivial order in $r$ requires that $F$ and $f_i$ satisfy the three differential equations
\begin{align}\label{StartDE}
 \mathbb{D}_{(i} f_{j)} +  F \mathbb{S}_{ij} = I_{ij},
\end{align}
where we have set
\begin{align}\label{Iij}
I_{ij} := \frac{1}{6} \overset{\mathtt{o}}{R}_{IKJL} \mathbb{B}^I_i \mathbb{B}^J_j r^K r^L  - \frac{1}{2c^2} W_I W_J \mathbb{R}^I_i \mathbb{R}^J_j .
\end{align}
Taking the trace and trace-free parts of these equations yields three equivalent equations:
\begin{align}
F &= - \frac{1}{2}\mathbb{D} \cdot f+\frac{1}{2} I , \label{eq:traceDE} \\
\mathbb{D}_{<i} f_{j>} &=  I_{<ij>}, \label{eq:tracefreeDE}
\end{align}
where $I := \mathbb{S}^{ij} I_{ij}$ is the trace, and $I_{<ij>} := I_{ij} - \frac{1}{2} \mathbb{S}_{ij} I$ the (symmetric) trace-free part of $I_{ij}$.  With the inhomogeneous ``source'' term $I_{ij}$ specified, equation~(\ref{eq:traceDE}) tells us that $F$ (the radial perturbation) is determined uniquely once $f_i$ (the angular perturbation) is known.  Thus, our focus will be on solving equation~(\ref{eq:tracefreeDE}) for $f_i$.  To do so, we simply expand $f_i$ as a sum of independent vector spherical harmonics with arbitrary coefficients, calculate $\mathbb{D}_{<i} f_{j>}$, decompose both $\mathbb{D}_{<i} f_{j>}$ and  $I_{<ij>}$ into independent tensor spherical harmonics, and then read off the required coefficients.  The result is:
\begin{align}
F =\,& \alpha_{I}(t) r^{I} +\frac{\kappa}{18} \overset{\mathtt{o}}{T}_{00} -\frac{1}{6c^2}  W^2 +\mathbb{F}, \label{eq:F} \\
f_i =\,&  \alpha_{I}(t) \mathbb{B}^{I}_{i} + \beta_{I}(t) \mathbb{R}^{I}_{i} +\frac{1}{4} \partial_{i} \mathbb{F} , \label{eq:fi}
\end{align}
where
\begin{equation}\label{script F}
\mathbb{F}:= Q^{IJ}\left( \frac{1}{c^2} W_I W_J + \frac{1}{3} \overset{\mathtt{o}}{\mathcal{E}}_{IJ} + \frac{\kappa}{6} \overset{\mathtt{o}}{T}_{IJ}\right)
\end{equation}
is a pure $\ell=2$ spherical harmonic.
Here $\alpha_I(t)$ and $\beta_I(t)$ are six arbitrary, time-dependent functions; $Q^{IJ} := r^I r^J - \frac{1}{3} \delta^{IJ}$ is trace-free and represents the five independent pure $l=2$ spherical harmonics; $W^2 := \delta^{IJ} W_I W_J$; and we have decomposed the Riemann tensor into the electric part of the Weyl tensor, $\mathcal{E}_{IJ} := C_{0I0J}$, the magnetic part of the Weyl tensor, $\mathcal{B}_{IJ} := \frac{1}{2} \epsilon_{I}^{\phantom{I}KL} C_{0JKL}$ (which we will need later), and the Ricci tensor, $\overset{\mathtt{o}}{R}_{ab}=\kappa(\overset{\mathtt{o}}{T}_{ab}-\frac{1}{2}\overset{\mathtt{o}}{T}\overset{\mathtt{o}}g_{ab})$, where we have dropped the superscript ``mat'' on the matter energy-momentum tensor, $\overset{\mathtt{o}}{T}_{ab}$.

It is instructive to take a moment to analyze this solution.  We begin with the homogeneous part, i.e., the solution to~(\ref{StartDE}) when $I_{ij}=0$.  This solution was discussed in detail in reference~\cite{Epp2009}.  It is given by the first term on the right-hand side of equation~(\ref{eq:F}) and the first two terms on the right-hand side of equation~(\ref{eq:fi}), i.e., $F =\alpha_{I}(t) r^{I}$ and $f_i =  \alpha_{I}(t) \mathbb{B}^{I}_{i} + \beta_{I}(t) \mathbb{R}^{I}_{i}$.  Using equation~(\ref{f^Idecomposition}), and the identity $\mathbb{S}^{ij}\mathbb{B}_i^I\mathbb{R}_j^J=-\epsilon^{IJ}_{\phantom{IJ}K}r^K$, this is equivalent to $f^I=\alpha^I(t) + \epsilon^{I}_{\phantom{I}JK}r^J\beta^K(t)$.  When multiplied by $r^3$, this is the spatial perturbation of the embedding of the observers' worldlines into the $X^I$ coordinate system.  Thus, $\alpha^I(t)$ and $\beta^I(t)$ clearly correspond to time-dependent translations and rotations of the RQF, respectively, and are precisely the same six degrees of freedom of rigid body motion we are familiar with in Newtonian space-time. They exist in this relativistic context because, on a two-sphere, the conformal Killing vector equation $\mathbb{D}_{<i} f_{j>}=0$ has precisely six independent solutions (the vectors $\mathbb{B}_I^i \partial_i$ and $\mathbb{R}_I^i \partial_i$ introduced above), which generate an action of the Lorentz group on the sphere.

It is important to note that: (1) While we are working at order $r^3$ here---see equation~({\ref{X^I=...}), i.e., the lowest order with a nontrivial particular solution, we find a homogeneous solution of the form discussed above for perturbations at both lower orders ($r$ and $r^2$) and higher orders, with no obvious reason this would change at arbitrarily high orders.  Thus, the general solution to the RQF rigidity equations has six arbitrary functions of time, $\alpha^I(t)$ and $\beta^I(t)$, at every order in $r$, or equivalently, six arbitrary functions of $t$ and $r$.  In other words, if we have ``nested'' RQFs, we are free to specify the ``Newtonian, $\ell=1$ vector spherical harmonic motion'' of each one {\it independently}.  (2) When we worked out various geometrical quantities (e.g., $\alpha_i$ and ${\mathcal P}_i$) at lowest order in the perturbation (order $r$), we noticed that $\alpha_I$ and $\beta_I$ were always paired with the Fermi frame proper acceleration, $A_I$, and proper rotation rate, $W_I$, in the combinations:
\begin{equation}\label{alpha-A and beta-W}
(A_I+r\ddot{\alpha}_I)\hspace{.5in}\text{and}\hspace{.5in} (W_I+\dot{\beta}_I).
\end{equation}
So at the lowest order, at least, for an RQF of given areal radius $r$, the perturbation generated by $\alpha_I$ (respectively, $\beta_I$) is equivalent to the corresponding acceleration (respectively, rotation rate) of the Fermi frame that the RQF is tied to.  Although we have not checked it at higher order, this is a natural result, and for simplicity's sake we will henceforth set the homogeneous solution at {\it all} orders in $r$ to zero,\footnote{It should be pointed out, however, that setting the homogeneous solution to the order $r^3$ perturbation to zero has no effect on any of our results.  One can show that the $\alpha_I$ and $\beta_I$ arising at this order of the perturbation do not appear in the results quoted below to the orders in $r$ to which they are displayed.} and take $A_I (t)$ and $W_I (t)$ as the six arbitrary, time-dependent degrees of freedom of the RQF.

Moving on to the particular solution, there are three points worth making in order to appreciate the physical significance of the various terms in equations~(\ref{eq:F}) and~(\ref{eq:fi}): (1) Positive mass-energy matter inside an RQF will ``warp'' the spatial slice spanning the round two-sphere boundary of the RQF (of areal radius $r$) in such a way that the proper radial distance to the ``centre'' of the RQF will be {\it larger} than $r$ (think of a standard funnel-shaped embedding diagram).  This explains the presence of the matter mass-energy term proportional to $\overset{\mathtt{o}}{T}_{00}$ in equation~(\ref{eq:F}).  (2) In reference~\cite{Epp2009} we considered a round sphere RQF of areal radius $r$ spinning with constant angular velocity, $\omega$, in flat spacetime.  We found that inertial observers outside the system would see a rotating, ``cigar''-shaped sphere with radial perturbation (at order $r^3$) given by $F = \frac{\omega^2}{c^2} (\cos^2 \theta - \frac{1}{2})$, where $\theta=0$ defines the rotation axis.  This $F$ corresponds to a radial contraction near the equator (to compensate for a circumferential Lorentz contraction) and a radial expansion near the poles (to maintain a pole-to-pole distance of $\pi r$ in spite of the radial contraction near the equator).  It is a simple exercise to check that equation~(\ref{eq:F}) (including the $W_I W_J$ term in $\mathbb{F}$) reduces to this expression for $F$ in this case.  This explains the presence of the rotation terms in equation~(\ref{eq:F}).  (3) The pure $\ell =2$ spherical harmonic term, $\mathbb{F}$, in equations~(\ref{eq:F}) and~(\ref{eq:fi}), also includes contributions from the electric part of the Weyl tensor ($\overset{\mathtt{o}}{\mathcal{E}}_{IJ}$), i.e., tidal forces, and spatial matter stresses ($\overset{\mathtt{o}}{T}_{IJ}$).  Both of these spatial curvature effects clearly need to be present in the coordinate perturbation required to achieve a round sphere RQF.

Having found the general solution to the RQF rigidity equations, we can now compute the intrinsic geometry of a generic RQF.  Recall from {\S}\ref{Intrinsic} that the two intrinsic geometrical degrees of freedom of an RQF can be encoded, in a coordinate independent manner, in the observers' proper acceleration tangential to ${\mathcal B}$. Computing the lapse and shift functions, $N$ and $u_i$, in the induced three-metric, equation~(\ref{eq:InducedMetric}), and substituting these into equation~(\ref{eq:ObserversAcceleration}), we find:
\begin{align}\label{eq:TangentialAcceleration}
&\alpha_i = r A_I \mathbb{B}^{I}_{i}  + r^2 \dot{W}_I \mathbb{R}^{I}_{i}
+ r^2 \left[ - \frac{1}{c^2} A_I A_J + W_I W_J + c^2 \overset{\mathtt{o}}{\mathcal{E}}_{IJ} -\frac{c^2\kappa}{2} \overset{\mathtt{o}}{T}_{IJ} \right] \mathbb{B}^{I}_{i} r^J + \mathcal{O}(r^3).
\end{align}
If we let $\mathbf{A}$ and $\mathbf{W}$ denote the vectors $A^I\partial_I$ and $W^I\partial_I$ in the Fermi spatial coordinate system $X^I$ (with $\partial_I:=\partial/\partial X^I$), then the contribution of the first term on the right-hand side to $\alpha^i\partial_i$ is the projection of $\mathbf{A}$ tangential to the RQF sphere, and the contribution of the second is $\mathbf{R}\times\dot{\mathbf{W}}$, where $\mathbf{R}$ is the radial vector from the origin of the coordinate system to observers on the RQF sphere.  Thus, these parts of $\alpha^i\partial_i$ are the direct result of the acceleration and rotation rate of the Fermi frame, to which the RQF is tied.  Of the other terms in equation~(\ref{eq:TangentialAcceleration}), the one involving the electric part of the Weyl tensor is interesting: it represents tidal forces, that is, tangential accelerations that the RQF observers must undergo in order to maintain rigidity, i.e., to compensate for the geodesic deviations they would otherwise experience in freefall. In the framework of gravitoelectromagnetism (GEM), we may follow Mashhoon in reference~\cite{Mashhoon} and define (at this order in $r$) the GEM electric and magnetic fields in the neighborhood of $\mathcal C$ as:
\begin{eqnarray}
E_I^\texttt{GEM}:= \,& c^2 \overset{\mathtt{o}}{\mathcal{E}}_{IJ}X^J, \label{GEM E Field} \\
B_I^\texttt{GEM}:= \,& -c^2 \overset{\mathtt{o}}{\mathcal{B}}_{IJ}X^J . \label{GEM B Field}
\end{eqnarray}
Then the part of $\alpha^i\partial_i$ that arises from the electric part of the Weyl tensor term in equation~(\ref{eq:TangentialAcceleration}) is easily seen to be $P^{IJ}E_I^\texttt{GEM}\partial_J$, i.e., the component of the GEM electric field (which is essentially acceleration in the GEM framework) tangential to the RQF sphere.

For completeness we also give the twist of the RQF congruence, computed using equation~(\ref{eq:nu}):
\begin{equation} \label{Twist}
\nu = - W_I r^I  + r \bigg[ \left( c \overset{\mathtt{o}}{\mathcal{B}}_{IJ} + \frac{2}{c^2} W_I A_J \right) r^I r^J - \frac{1}{c^2} \delta^{IJ} A_I W_J \bigg] + \mathcal{O}(r^2).
\end{equation}
The twist measures the proper rotation rate of observers' spatial dyads relative to inertial gyroscopes.  As one would expect, at lowest order the twist is the (negative) of the radial component of the rotation rate of the Fermi frame.  At next order in $r$, the most interesting term is the one involving the magnetic part of the Weyl tensor; it is interesting because rotation is believed to be one of the sources of ${\cal B}_{IJ}$~\cite{Bonnor1995}. In terms of Mashhoon's definition of the related GEM magnetic field in equation~(\ref{GEM B Field}), the part of $\nu$ in question is easily seen to be $-r^I B_I^\texttt{GEM}/c$, which, in the GEM framework, is the radial component of the rotation vector, $- B_I^\texttt{GEM}/c$~\cite{Mashhoon}.

We mentioned in {\S}\ref{Intrinsic} that the RQF intrinsic geometrical degrees of freedom are essentially encoded in $\alpha_i$, discussed above, but that we are also free to specify the twist on one cross section of $\mathcal B$. To see this at lowest order, notice that if we specify the $\ell=1$ component of $\alpha_i$, i.e., $\alpha_i = r A_I \mathbb{B}^{I}_{i}  + r^2 \dot{W}_I \mathbb{R}^{I}_{i}$, then this determines $A_I(t)$ and $\dot{W}_I(t)$.  To know $W_I(t)$, i.e., the full six degrees of freedom, we must also specify $W_I(0)$ at some initial time $t=0$, which we do when we specify the $\ell=1$ component of $\nu$ on an initial slice of $\mathcal B$.

Let us now turn our attention to the extrinsic geometrical quantities associated with an RQF, beginning with the momentum surface density, ${\mathcal P}_a$, appearing in the energy conservation law in equation~(\ref{eq:SimpleConservationEq}). Starting at equation~(\ref{SurfaceSEMcomponents}), we find:
\begin{equation}\label{eq:SurfaceMomentumAW}
{\mathcal P}_i = \frac{1}{c\kappa} r W_I \mathbb{R}^{I}_{i} + r^2 \bigg[ - \frac{1}{c \kappa}\left( c \overset{\mathtt{o}}{\mathcal{B}}_{IJ} + \frac{2}{c^2} W_I A_J \right) \mathbb{R}^{I}_{i} r^J - \frac{1}{2} \overset{\mathtt{o}}{T}_{0I} \mathbb{B}^{I}_{i}\bigg] + \mathcal{O}(r^3).
\end{equation}
The last term on the right-hand side is clearly associated with the matter momentum density projected tangentially to the RQF sphere, and so makes sense intuitively.  However, from equation~(\ref{SurfaceSEMcomponents}) we recall that ${\mathcal P}^a$ can also be interpreted in terms of the precession rate of inertial gyroscopes (projected tangentially to $\mathcal B$ and rotated by 90 degrees).  Apart from a common factor of $-r/c\kappa$, the first three terms in equation~(\ref{Twist}) are identical to the first three terms in equation~(\ref{eq:SurfaceMomentumAW}), except in the former case we have the contraction of a rotation vector with $r^I$ (projection normal to the RQF sphere), and in the latter case we have the contraction of the {\it same} rotation vector with $\mathbb{R}^I_i$ (projection tangential to the RQF sphere and rotated by 90 degrees).

We now proceed to calculate both the matter and geometrical energy fluxes appearing on the right-hand side of our energy conservation law in equation~(\ref{eq:SimpleConservationEq}).  Note that $d{\mathcal B}/c=r^2 \, d\mathbb{S}\,dt\,N$, where $d\mathbb{S}$ is the area element on a unit round sphere, and $N$ is the lapse function associated with our choice of time foliation of $\mathcal B$.  So we are really interested in the matter and geometrical energy fluxes {\it times} the lapse function.  A straightforward but tedious calculation reveals (recall that we have dropped the superscript ``mat'' on the matter energy-momentum tensor, $T_{ab}$):
\begin{align}
N \left( n^a u^b T_{ab}\right) &=  -r^I \overset{\mathtt{o}}{S}_I + r \left( \frac{1}{3} \frac{\partial \overset{\mathtt{o}}{\rho}}{\partial t} + \frac{1}{3 c^2} \overset{\mathtt{o}}{S}_I A^I + \Psi_{\texttt{mat}} \right) + \mathcal{O}(r^2)  \label{ndotS} , \\
N\left(-\alpha \cdot \mathcal{P} \right)&=   -\frac{c^2}{8\pi G} \epsilon_{IJK} r^I A^J W^K
  + r \left( -\frac{1}{3 c^2} \overset{\mathtt{o}}{S}_I A^I  -\frac{1}{3}\frac{c^2}{8\pi G}  \frac{\partial W^2}{\partial t} + \Psi_{\texttt{geo}} \right) + \mathcal{O}(r^2)   \label{adotP},
\end{align}
where $\overset{\mathtt{o}}{\rho} := \overset{\mathtt{o}}{T}_{00} $ is the matter energy density (energy per unit volume) evaluated on $\mathcal C$, i.e., at the ``centre'' of the sphere; $\overset{\mathtt{o}}{S}_I := -c \overset{\mathtt{o}}{T}_{0I}$ is the matter energy flux (power per unit area) in the $X^I$ direction, evaluated at the ``centre'' of the sphere; and
\begin{align}
& \Psi_{\texttt{mat}} := Q^{IJ} \left( -\overset{\mathtt{o}}{S}_{I;J} - \frac{1}{c^2} \overset{\mathtt{o}}{S}_I A_J - \overset{\mathtt{o}}{T}_{IK} \epsilon^{K}_{\phantom{K}JL} W^L \right), \label{psi_mat} \\
& \Psi_{\texttt{geo}} := Q^{IJ} \left( \frac{1}{2c^2} \overset{\mathtt{o}}{S}_I A_J - \frac{1}{2} \overset{\mathtt{o}}{T}_{IK} \epsilon^{K}_{\phantom{K}JL} W^L + \frac{c^2}{16\pi G} \frac{\partial}{\partial t} (W_I W_J)\right.  \nonumber \\
& \hspace{30mm} \left. + \frac{1}{\kappa} \epsilon_{J}^{\phantom{J}KL} \left[\frac{2}{c^4} A_K W_L A_I - \frac{1}{c} A_K \overset{\mathtt{o}}{\mathcal{B}}_{LI} + W_K \overset{\mathtt{o}}{\mathcal{E}}_{LI} \right] \right) \label{psi_geo}.
\end{align}
Equations~(\ref{ndotS}) and~(\ref{adotP}) represent the {\it negative} of outgoing fluxes, and according to equation~(\ref{eq:SimpleConservationEq}), if we multiply these by $r^2 \, d\mathbb{S}\,dt$, add them, and integrate over the angles of the sphere, and time, we will get the change in the total energy of the RQF (matter plus gravitational) between initial and final time slices (${\mathcal S}_i$ and ${\mathcal S}_f$) of $\mathcal B$.

Let us try to understand the physical significance of the various individual flux terms in these equations.  First, recall that $Q^{IJ} := r^I r^J - \frac{1}{3} \delta^{IJ}$ is trace-free and represents the five independent pure $\ell=2$ spherical harmonics, so $\Psi_{\texttt{mat}}$ and $\Psi_{\texttt{geo}}$ both vanish when integrated over the angles.  These are similar in character to the near-field energy fluxes in an electromagnetically radiating system, in that there is energy flowing inwards and outwards, with no net flux.  For example, in $\Psi_{\texttt{geo}}$, there are cross products of acceleration with the magnetic part of the Weyl tensor, and rotation with the electric part of the Weyl tensor.  Considering the close relationship between acceleration and electric-like effects of gravity, and rotation and magnetic-like effects of gravity, which we will see more of below, these terms are similar in spirit to a gravitational analogue of the electromagnetic Poynting vector.  However, as interesting as they may be, insofar as they do not contribute to the integrated flux, we will leave a detailed analysis of $\Psi_{\texttt{mat}}$ and $\Psi_{\texttt{geo}}$ for future work.

The first term on the right-hand side of equation~(\ref{ndotS}) is the inward radial projection of the matter energy flux evaluated on the RQF sphere, to lowest order in $r$; the latter is constant, and equal to its value at the centre of the sphere, i.e., $\overset{\mathtt{o}}{S}_I$.  The result is obviously a pure $\ell=1$ spherical harmonic that integrates to zero over the angles.  For example, if $\overset{\mathtt{o}}{S}_I$ is in the $z$-direction, then $-r^I \overset{\mathtt{o}}{S}_I$ will be proportional to $-\cos\theta$, and the fact that it integrates to zero just says that, to lowest order in $r$, whatever matter flux enters through the bottom half of the sphere must leave the top half of the sphere.

The corresponding lowest order term in the geometrical energy flux---the first term on the right-hand side of equation~(\ref{adotP}), is similarly a pure $\ell =1$ spherical harmonic that integrates to zero over the angles.  However, its interpretation is worth discussing.  Comparing the lowest order matter and geometrical energy fluxes we have the correspondence: $\overset{\mathtt{o}}{S}_I$ (matter flux) $\leftrightarrow \frac{c^2}{8\pi G} \epsilon_{IJK} A^J W^K$ (geometrical flux). Thus, at lowest order, the geometrical energy flux is proportional to the cross product of the Fermi frame acceleration and rotation rate.  It exists even in flat spacetime, and can be motivated through the equivalence principle as follows.

We imagine an RQF in flat spacetime undergoing arbitrary, but slow motion, time-dependent acceleration and rotation.  Retaining terms only linear in the acceleration and rotation, and setting curvature and matter terms to zero, the observers' tangential acceleration can be read off from equation~(\ref{eq:TangentialAcceleration}): $\alpha_i = rA_I(t)\mathbb{B}_i^I+r^2\dot{W}_I(t)\mathbb{R}_i^I$.  We now consider a spacetime in general, linearized gravity, with line element:\cite{Mashhoon}
\begin{equation}\label{GEM Line Element}
ds^2 = -c^2\left(1+2\frac{\Phi}{c^2}\right)\,dT^2+\frac{4}{c}{\mathcal A}_I\,dX^I\,dT+ \left(1-2\frac{\Phi}{c^2}\right)\delta_{IJ}\,dX^I\,dX^J,
\end{equation}
where, in the Newtonian limit, $\Phi$ reduces to the Newtonian gravitational potential, and ${\mathcal A}_I$ is a vector potential associated with rotation of the spacetime.  Comparing with equations~(\ref{T=t}) to~(\ref{f^Idecomposition}), we now embed RQF observers who are `at rest' in this spacetime via the coordinate transformation:
\begin{equation}\label{GEM Embedding}
T=t\hspace{.5 in}\text{and}\hspace{.5 in}X^I=r\left(1+F\right)r^I.
\end{equation}
A quick calculation shows that the RQF rigidity equations are satisfied when we choose $F=\Phi/c^2$.  Computing $N$ and $u_i$, and substituting these into equation~(\ref{eq:ObserversAcceleration}), we find that observers `at rest' in this linearized gravitational field experience a tangential gravitational force per unit mass given by $-\alpha_i=-\partial_i\Phi-2r\dot{\mathcal A}_I\mathbb{B}_i^I/c$.  In the spirit of the equivalence principle, we now ask, ``Can we find gravitational potentials $\Phi$ and ${\mathcal A}_I$ such that RQF observers `at rest' in this gravitational field experience the {\it same} tangential gravitational force per unit mass as they do while accelerating and tumbling in flat spacetime, and so cannot distinguish between these two situations?''  In other words, we wish to equate $-\alpha_i=-rA_I(t)\mathbb{B}_i^I-r^2\dot{W}_I(t)\mathbb{R}_i^I$ (inertial gravitational field in flat spacetime) with $-\alpha_i=-\partial_i\Phi-2r\dot{\mathcal A}_I\mathbb{B}_i^I/c$ (gravitational force per unit mass associated with remaining `at rest' in a linearized gravitational field).  Equating (the negative of) these two accelerations results in the required gravitational potentials: $\Phi=A_I(t)x^I$ and ${\mathcal A}_I=c\,\epsilon_{IJK}x^J W^K (t)/2$, where $x^I := rr^I$. Now we ask, ``Is there a gravitational energy flux associated with these gravitational potentials?''  According to the gravitoelectromagnetic (GEM) interpretation of linearized gravity, these gravitational potentials are associated with gravitoelectric and gravitomagnetic vector fields.  Using the formulas in reference~\cite{Mashhoon} we find (in obvious boldface vector notation): ${\mathbf E}^\texttt{GEM}={\mathbf A}(t)+\frac{1}{4}{\mathbf r}\times\dot{{\mathbf W}}(t)$ and ${\mathbf B}^\texttt{GEM}=c{\mathbf W}(t)$.  Within this same interpretation, there `ought' to be an associated GEM Poynting vector, ${\mathbf S}^\texttt{GEM}$, proportional to ${\mathbf E}^\texttt{GEM}\times {\mathbf B}^\texttt{GEM}$~\cite{Mashhoon}.  We can determine this proportionality constant by comparing our expression for ${\mathbf E}^\texttt{GEM}\times {\mathbf B}^\texttt{GEM}$ with the lowest order result in equation~(\ref{adotP}); we find:\footnote{Notice that the proportionality constant here (determined using our coordinate invariant RQF approach) differs from that obtained in reference~\cite{Mashhoon} (determined using a coordinate-dependent pseudotensor approach).}
\begin{equation}\label{GEMPoynting}
{\mathbf S}^\texttt{GEM}=\frac{c}{8\pi G} {\mathbf E}^\texttt{GEM}\times {\mathbf B}^\texttt{GEM} = \frac{c^2}{8\pi G}\left[{\mathbf A}\times{\mathbf W}+\frac{1}{4}\left({\mathbf r}\times\dot{{\mathbf W}}\right)\times{\mathbf W}\right].
\end{equation}
Thus we see some justification in the above argument, based on the equivalence principle, for both the existence of the geometrical energy flux even in flat spacetime, and its interpretation as a gravitational energy flux.

Related to the previous discussion, the geometrical energy flux in equation~(\ref{adotP}) contains a term proportional to the time derivative of $W^2$, which does not vanish upon integration over the angles.  We will see below, when we evaluate the left-hand side of equation~(\ref{eq:SimpleConservationEq}), that there is a correctly matching $W^2$ term contributing to the energy of the RQF---see equation~(\ref{LHSintergral}).  Two comments on this rotational contribution to the RQF energy are worth making. (1) It can be accounted for using our GEM Poynting vector in equation~(\ref{GEMPoynting}), but not perfectly.  When we compute the radial component of the term proportional to $({\mathbf r}\times\dot{{\mathbf W}})\times{\mathbf W}$, we find a part that integrates to zero over the angles, and a part that does not.  The latter is proportional to the time derivative of $W^2$, but the numerical factor in the proportionality constant does not match what we have in equation~(\ref{adotP}).  So it agrees in spirit, but not in detail. However, this is not unexpected, since the GEM calculation is in the context of linearized gravity, and nonlinear effects could very well contribute a term of this form.  We emphasize that equations~(\ref{adotP}) (and \ref{ndotS}) are {\it exact} (to the displayed order in $r$), accounting fully for the nonlinearity of general relativity.  The GEM calculation, on the other hand, is approximate, and used here for motivational purpose only.  (2) Inspection of the sign in equation~(\ref{adotP}), or~(\ref{LHSintergral}) below, reveals that the rotational contribution to the RQF energy is {\it negative}: if $W^2$ increases, the RQF energy decreases.  One possibly plausible explanation for the sign (and a second argument for the very existence of this rotational energy) is that the (``unreferenced'') quasilocal energy density, $\mathcal E$, contains a {\it negative} vacuum contribution.  Looking ahead to equation~(\ref{eq:SurfaceEnergyAW}), the vacuum energy surface density is the first term on the right-hand side, ${\mathcal E}^\texttt{vac}:=-2/\kappa r$, which integrates to $E^\texttt{vac}=-c^4 r/G$ over the surface of the RQF sphere. As mentioned in a footnote in {\S}\ref{Extrinsic}, this vacuum energy is irrelevant when computing {\it changes} in energy, but it may, after all, be {\it indirectly} relevant.  If this energy is actually present ``inside'' an RQF, even when the RQF is in flat spacetime and not rotating, then after spinning up the RQF observers, perhaps the RQF observers are rotating relative to this vacuum energy, and as such `ought' to perceive this as a kind of rotational kinetic energy.  Since the vacuum energy is negative, presumably any moment of inertia that might be associated with $E^\texttt{vac}$ would also be negative, and hence the negative rotational kinetic energy.  Turning the argument around, we might say that the existence of a negative rotational kinetic energy indirectly implies the existence of a negative vacuum energy.

We now turn to what might be considered the main, and perhaps most interesting, flux terms in equations~(\ref{ndotS}) and~(\ref{adotP}).  These are the matter energy flux terms: $r \left[ \frac{1}{3} \frac{\partial \overset{\mathtt{o}}{\rho}}{\partial t} + \frac{1}{3 c^2} \overset{\mathtt{o}}{S}_I A^I \right]$, and the geometrical energy flux term: $r \left[ -\frac{1}{3 c^2} \overset{\mathtt{o}}{S}_I A^I  \right]$.  Integrating the first matter energy flux term over the RQF sphere, i.e., multiplying by $4\pi r^2$, gives $V\partial \overset{\mathtt{o}}{\rho}/\partial t$ (where $V=4\pi r^3/3$ is the proper volume of the RQF sphere), i.e., the proper time rate of change of the matter energy inside the RQF, to lowest order in $r$.  This is an expected result, correctly matched by the corresponding matter energy term on the right-hand side of equation~(\ref{LHSintergral}) below.  However, there is a second matter energy flux term, which couples the `standard' matter energy flux ($\overset{\mathtt{o}}{S}_I$) with the acceleration of the (rigid quasilocal) frame.  This term does not, in general, integrate to zero, and so if we used only $N\left( n^a u^b T_{ab} \right)$ to evaluate the change in matter energy of an accelerating system, we would get the wrong answer.  However, being of the opposite sign, the geometrical energy flux term is exactly what is required to cancel this extra acceleration-induced flux term, resulting in the correct answer.  To see the physical significance of this cancellation process, and the necessity of the geometrical energy flux term, we will now construct an apparent paradox in special relativity and resolve it using these RQF results.

Consider a cylinder of length $L$ and cross sectional area $A$, whose axis is parallel to the $z$-axis of an inertial reference frame in flat spacetime, with Minkowski coordinates $(t,x,y,z)$.  The cylinder sits in a uniform electric field, $\vec{E} = \hat{x}E$, parallel to the $x$-axis, and contains electromagnetic energy $E^2/8\pi$ times the volume of the cylinder, $AL$.  We will now subject the cylinder to a constant proper acceleration in the positive $z$ direction in such a way that it is an RQF, and ask how the electromagnetic energy in the cylinder changes with time. We will compute this change using two methods: (1) the change in the volume energy density (times the volume), and (2) the net Poynting flux integrated over the surface.  The paradox is that these two methods will give different answers.  The resolution of this apparent paradox will involve the RQF geometrical energy flux, which we will interpret as a bona fide gravitational effect. While it can be understood superficially in the context of special relativity, its deeper explanation lies in general relativity.

First, we need to accelerate the cylinder in such a way that it is an RQF.  Let the bottom of the cylinder have constant proper acceleration, $a$.  It is well known that, in order for the length of the cylinder to remain constant for co-moving (RQF) observers (the requirement for an RQF), the top of the cylinder must experience {\it less} proper acceleration, namely, $a^\prime=a/(1+aL/c^2)$.  This simple fact is usually called Bell's spaceship paradox (and is not the paradox we are concerned with here). Since the dimensions of the cross sections of the cylinder are not affected by this acceleration, we thus have an RQF. There are two important facts to note about this RQF. (1) Proper time moves at different relative rates for observers at the bottom and top of the cylinder. If, between two simultaneities for the RQF observers, a proper time $\Delta\tau$ elapses for observers at the bottom, a {\it greater} proper time, $\Delta\tau^\prime =(1+aL/c^2)\Delta\tau$ elapses for observers at the top. (2) While the relative velocity, $v$, between RQF observers and the inertial reference frame is of course changing (increasing), on any given RQF simultaneity all RQF observers see the {\it same} instantaneous relative velocity.  So we can use the relative velocity $v$ to label the RQF simultaneities.

Next, let us consider the electromagnetic field the RQF observers see.  Since they are moving perpendicular to an electric field, they will see, in addition to a stronger electric field, also a magnetic field: $\vec{E}^\prime = \hat{x} \gamma E $ and $\vec{B}^\prime = - \hat{y}\beta \gamma E $, where $\beta=v/c$ and $\gamma=1/\sqrt{1-\beta^2}$.  Note that since these fields depend only on $v$, all RQF observers on any given RQF simultaneity will instantaneously see the {\it same} electric and magnetic fields.  They will thus see the same Poynting vector, $\vec{S}^\prime = \frac{c}{4\pi}\vec{E}^\prime \times \vec{B}^\prime = - \hat{z}\frac{c}{4 \pi} \beta \gamma^2 E^2 $, and the same volume energy density, $u^\prime = \frac{1}{8 \pi} (E^{\prime 2} + B^{\prime 2}) = \frac{1}{8 \pi} (1 + \beta^2) \gamma^2 E^2$.  Note that, according to this expression for $u^\prime$, the total electromagnetic energy inside the cylinder is clearly increasing with time.  The question is, ``What is the mechanism responsible for this increase?''

Now we will calculate the change in the electromagnetic energy in the cylinder as seen by the RQF observers.  The natural way to parameterize this change is to consider the change in the electromagnetic energy between a pair of infinitesimally separated RQF simultaneities, labeled by, say, proper time $\tau$ and $\tau+\Delta\tau$ as experienced by observers at the bottom of the cylinder.  (A bit of thought shows that it does not matter whose proper time we use to parametrize the simultaneities.)  As mentioned above, we can then calculate this change using two different methods: (1) The volume energy density method, and (2) the Poynting flux method.  For method (1), note that the proper volume of the RQF is constant (by the nature of it being an RQF) and equal to $AL$, which we will denote as $V$.  Thus, $\Delta E^{(1)} = V(du^\prime /d\tau)\Delta\tau$.  Now $u^\prime = \frac{1}{8 \pi} (1 + \beta^2) \gamma^2 E^2$ depends only on $v$, so to calculate $du^\prime /d\tau$ we need to know $dv /d\tau$, which is equal to $a/\gamma^2$ for an observer experiencing constant proper acceleration, $a$.  A simple calculation then yields:
\begin{equation}\label{E(1)}
\Delta E^{(1)}=\frac{1}{2\pi}\beta\gamma^2 E^2 V\frac{a}{c}\Delta\tau.
\end{equation}
This result is analogous to using the first of the two matter energy flux terms discussed above: $r \left[ \frac{1}{3} \frac{\partial \overset{\mathtt{o}}{\rho}}{\partial t} \right]$, and is the correct answer.

For method (2)---the Poynting flux method, recall that all observers on a given RQF simultaneity (in particular, those at the bottom and top of the cylinder) see the {\it same} Poynting flux, $\vec{S}^\prime = - \hat{z}\frac{c}{4 \pi} \beta \gamma^2 E^2 $.  On first thought this may seem to be a problem, since wouldn't an equal flux flowing in through the top and out through the bottom mean no net change in the electromagnetic energy?  What saves us is the fact that, due to the differing proper accelerations, proper time flows more quickly at the top of the cylinder relative to the bottom.  As noted above, if---between two RQF simultaneities---a proper time $\Delta\tau$ elapses at the bottom, a greater proper time, $\Delta\tau^\prime =(1+aL/c^2)\Delta\tau$, elapses at the top.  Thus, between two RQF simultaneities, more proper time elapses at the top, allowing more energy to enter through that surface than exits through the bottom.  This is apparently the mechanism explaining {\it how} the electromagnetic energy inside the cylinder increases with time.  We say ``apparently'' because it doesn't quite give the right answer.  With the magnitude of the Poynting vector given by $\frac{c}{4 \pi} \beta \gamma^2 E^2$, we have $\Delta E^{(2)} = A \frac{c}{4 \pi} \beta \gamma^2 E^2 \left(\Delta\tau^\prime - \Delta\tau\right)$, and so:
\begin{equation}\label{E(2)}
\Delta E^{(2)}=\frac{1}{4\pi}\beta\gamma^2 E^2 V\frac{a}{c}\Delta\tau.
\end{equation}
Clearly, this accounts for only half of the correct answer: $\Delta E^{(2)}=\Delta E^{(1)}/2$.  This approach is analogous to using {\it both} of the matter energy flux terms discussed above: $r \left[ \frac{1}{3} \frac{\partial \overset{\mathtt{o}}{\rho}}{\partial t} + \frac{1}{3 c^2} \overset{\mathtt{o}}{S}_I A^I \right]$.  In fact, if we replace the cylinder in this example with a round sphere RQF, one can show that, numerically, $\frac{1}{3 c^2} \overset{\mathtt{o}}{S}_I A^I = - \frac{1}{2} (\frac{1}{3} \frac{\partial \overset{\mathtt{o}}{\rho}}{\partial t})$ so that $\frac{1}{3} \frac{\partial \overset{\mathtt{o}}{\rho}}{\partial t} + \frac{1}{3 c^2} \overset{\mathtt{o}}{S}_I A^I = \frac{1}{2} (\frac{1}{3} \frac{\partial \overset{\mathtt{o}}{\rho}}{\partial t})$, which explains the factor of one-half.
Furthermore, using these results, and the fact that $N=1+\frac{1}{c^2}rr^I A_I + \mathcal{O}(r^2)$, it is easy to show that
$n^a u^b T_{ab}$ (the flux without the lapse function in front) is equal to
$$
-r^I \overset{\mathtt{o}}{S}_I + r \left[ \frac{1}{3} \frac{\partial \overset{\mathtt{o}}{\rho}}{\partial t} + \frac{2}{3 c^2} \overset{\mathtt{o}}{S}_I A^I  \right]+ \mathcal{O}(r^2) = -r^I \overset{\mathtt{o}}{S}_I + \mathcal{O}(r^2)
$$
(ignoring $\Psi_{\texttt{mat}}$).  In other words, $-n^a u^b T_{ab}$ is analogous to $\vec{S}^\prime$ in the cylinder example.  Multiplying $n^a u^b T_{ab}$ by the lapse function, $N$, is equivalent to taking into account the difference in proper times, $\Delta\tau^\prime =(1+aL/c^2)\Delta\tau$, in the cylinder example.  In short, the Poynting flux method is analogous to using $N\left( n^a u^b T_{ab}\right)$ to compute the change in electromagnetic energy, which one might {\it think} is the correct thing to do, but it is not.  This is the paradox.

This paradox is resolved by including the geometrical energy flux term, $r \left[ -\frac{1}{3 c^2} \overset{\mathtt{o}}{S}_I A^I  \right]$, coming from $\alpha\cdot {\mathcal P}$.  There are two senses in which this geometrical energy flux can be thought of as a bona fide gravitational energy flux. (1) The mechanism behind the Poynting flux method here relies entirely on the fact that in an accelerating frame, proper time flows more quickly at the top relative to the bottom.  According to the equivalence principle, this situation is in essence the same as gravity.  So in our special relativity calculation above, we are really encroaching on the domain of gravity.  But to do it properly, we must use general relativity, not accelerating frames in flat spacetime.  The difference amounts to adding the geometrical flux term, which is thus seen to be a bona fide gravitational effect; so being in the context of an energy flux, it must be a {\it gravitational} energy flux.  It is amusing to compare this situation with the gravitational deflection of light.  It is well known that using the principle of equivalence to calculate the deflection of light gives exactly one-half of the correct result calculated using general relativity~\cite{Comer1978}. (2) In the ``real world'' we have $G_{ab}=\kappa T_{ab}$.  In special relativity with an electromagnetic field, on the other hand, we have $T_{ab}\not= 0$, but $G_{ab}=0$. In going from special to general relativity we allow the electromagnetic field to curve the geometry.  It is not unreasonable to imagine that an electromagnetically curved geometry gives rise to gravitational (curvature) effects that account for at least some of the effects of the electromagnetic field. In fact, a more detailed analysis~\cite{EMM2012} starting with the metric for a homogeneous electromagnetic field in general relativity~\cite{Stephani} reveals that this is exactly what is happening---the geometrical energy flux {\it is} a gravitational energy flux. Precisely half of the energy entering the cylinder is due to a traditional matter energy flux (Poynting vector), and the other half is due to a novel gravitational energy flux associated with the spacetime curvature created by the electromagnetic field. In general relativity all forms of energy (e.g., electromagnetic and gravitational) are equivalent, and the sum yields the correct total energy.

Having discussed the matter and gravitational flux terms appearing on the right-hand side of our energy conservation law in equation~(\ref{eq:SimpleConservationEq}), we now turn our attention to the left-hand side of this equation, both for its own sake, and to provide a useful check of the (integrated) flux expressions in equations~(\ref{ndotS}) and~(\ref{adotP}). A short calculation reveals that the quasilocal energy surface density is given by
\begin{align}\label{eq:SurfaceEnergyAW}
&\mathcal{E} = - \frac{2}{\kappa r} - \frac{r}{\kappa} \left[ \left( \frac{3}{c^2} W_I W_J + \delta^{KL} \overset{\mathtt{o}}{R}_{IKJL} \right) r^I r^J -\frac{1}{2} \delta^{IJ} \delta^{KL} \overset{\mathtt{o}}{R}_{IKJL} \right].
\end{align}
The first term is a negative vacuum energy, discussed earlier. Using the fact that $\alpha\,d{\mathcal S}=r^2 \, d\mathbb{S}$, the result
$\beta_i = u_i = r^2 W_I \mathbb{R}^{I}_{i} + \mathcal{O}(r^3)$,
and the earlier result for ${\mathcal P}_i$, a straightforward calculation yields
\begin{align}\label{LHSintergral}
\int\limits_{{\mathcal S}_f-{\mathcal S}_i} d{\mathcal S}\, \alpha \left({\mathcal E} + \beta \cdot {\mathcal P} \right) =
\left[ \frac{4 \pi r^3}{3} \overset{\mathtt{o}}{\rho} - r^3 \frac{c^2}{6G} W^2 \right]_{t_i}^{t_f},
\end{align}
consistent with the integral of the matter and gravitational fluxes discussed earlier.  It is worth noting that both $\mathcal E$ and  $\beta \cdot {\mathcal P}$ contribute to give the correct numerical factor for the $W^2$ rotational kinetic energy term on the right-hand side.

To further strengthen the evidence that RQFs can be constructed in generic spacetimes, and to further explore the interpretation of the geometrical energy flux as a gravitational energy flux, we have carried out calculations to two higher orders in powers of $r$.  To make the calculations tractable we have have gone to a nonrotating Fermi normal coordinate system centered on a geodesic, i.e., $A_I = 0 = W_I$.  We have also turned off the matter sources, $R_{ab} = 0$, leaving only the electric and magnetic parts of the Weyl tensor.  With the matter sources off, the matter energy flux will vanish ($n^a u^b T_{ab}=0$), leaving only the geometrical energy flux.  At order $r^2$ (order $r^4$ when integrated over the RQF sphere), i.e., one order higher than in equation~(\ref{adotP}), the (outward) geometrical energy flux is found to be
\begin{align}
N\left(\alpha \cdot \mathcal{P}\right) =   \frac{c}{8\pi G}   \,\epsilon^{IJK}\,r_I\, E_J^\texttt{GEM} B_K^\texttt{GEM}
\end{align}
where $E_I^\texttt{GEM}$ and $B_I^\texttt{GEM}$ are the Weyl tensor-type GEM fields defined in equations~(\ref{GEM E Field}) and~(\ref{GEM B Field}).  At this order, this result is in agreement with Mashhoon's definition of a GEM Poynting vector ~\cite{Mashhoon}, which again adds more weight to the interpretation of the geometrical energy flux as a gravitational energy flux.  Note that the flux at this order is composed only of pure $\ell=3$ and $\ell=1$ spherical harmonics, and thus integrates to zero over the RQF sphere.  It represents a ``near field-like'' energy flux, flowing into and out of the RQF sphere with no net energy flow.

At the next order in $r$, the flux is composed of $\ell=4$, $\ell=2$ and $\ell=0$ parts.  For simplicity, we give only the integrated (outward) flux (i.e., the $\ell=0$ part):
\begin{align}\label{GravEnergyAtFifthOrder}
&\frac{1}{c}\int_{ \Delta {\mathcal B}} d{\mathcal B} \, \alpha \cdot \mathcal{P} = \left[ \frac{1}{60}\frac{c^4}{G} r^5  \left(\overset{\mathtt{o}}{\mathcal{E}} {}^2 - 2 \overset{\mathtt{o}}{\mathcal{B}} {}^2\right) \right]_{t_i}^{t_f},
\end{align}
where $\overset{\mathtt{o}}{\mathcal{E}} {}^2 = \overset{\mathtt{o}}{\mathcal{E}} {}_{IJ} \overset{\mathtt{o}}{\mathcal{E}} {}^{IJ}$ and $\overset{\mathtt{o}}{\mathcal{B}} {}^2 = \overset{\mathtt{o}}{\mathcal{B}} {}_{IJ} \overset{\mathtt{o}}{\mathcal{B}} {}^{IJ}$.

At first sight, the relative factor of $-2$ between $\overset{\mathtt{o}}{\mathcal{E}} {}^2$ and $\overset{\mathtt{o}}{\mathcal{B}} {}^2$ may seem troubling, both in magnitude and in sign.  For example, based on both the energy density in electromagnetism, and the ``0000'' component of the Bel-Robinson tensor, one might have expected an expression proportional to $(\overset{\mathtt{o}}{\mathcal{E}} {}^2 + \overset{\mathtt{o}}{\mathcal{B}} {}^2)$.  However, it is actually not clear {\it what} to expect.  For example, in reference~\cite{EMM2013} we discuss how the left-hand side of equation~(\ref{eq:SimpleConservationEq}) is exactly analogous to the {\it covariant} definition of electromagnetic energy given in equation~(16.44) of Jackson~\cite{Jackson3rdEdition}. In the case of a purely electrostatic system, viewed by a moving observer, Jackson shows that the correct integrand for the electromagnetic energy is proportional not to $(\mathbf{E}^2+\mathbf{B}^2 )$, but rather $(\mathbf{E}^2-\mathbf{B}^2 )$ (see his equation~(16.46)), and in the case of a non-purely electrostatic system the integrand is more complicated. So the fact that our expression for gravitational energy at order $r^5$ in equation~(\ref{GravEnergyAtFifthOrder}) is not proportional to $(\overset{\mathtt{o}}{\mathcal{E}} {}^2 + \overset{\mathtt{o}}{\mathcal{B}} {}^2)$ is perhaps not troublesome at all. Understanding this result more fully is an open problem we hope to pursue in the future. In any case, the main point we would like to make is that the RQF approach gives an extremely simple, operational definition for gravitational energy flux: $\alpha\cdot {\mathcal P}$.  When we expand it in powers of $r$ we get curvature tensor expressions that strongly suggest we are dealing with a bona fide gravitational energy flux, but the terms in the series will clearly get increasingly more complicated at higher orders in $r$.  Perhaps this is simply because ``curvature tensor expressions'' is not the correct language for gravitational energy. The RQF approach suggests that the correct language is a coupling between intrinsic and extrinsic curvature of the system boundary in the form $\alpha\cdot {\mathcal P}$.  This is a simple, exact, operational definition that is physically well-motivated.

\section{Conclusion}

In this chapter we have provided strong evidence that the notion of a {\it rigid quasilocal frame} can be extended from flat to curved spacetime.  We have presented a completely general solution of the RQF rigidity equations in an expansion in areal radius, based on Fermi normal coordinates, up to third order.  In the case of vanishing acceleration, rotation, and sources we were able to push this solution up to fifth order.  While the amount of algebra involved in such calculations grows very quickly, there do not appear to be any technical obstructions to extending these solutions to any order.  In other words, for all practical purposes it seems that the RQF equations can be satisfied in an arbitrary curved spacetime, at least out to the radius at which acceleration horizons form.

One of the motivations for introducing RQFs is to provide a new approach to the problem of motion, in particular, to allow the motion of a system to be analyzed in terms of natural, well-defined fluxes passing through the system boundary. Here we have seen that, within the context of both flat and curved spacetimes, the time component of the momentum constraint equation of general relativity, along with the notion of an RQF, give rise to a natural definition for the flux of gravitational energy, namely, $\alpha \cdot \mathcal{P}$.  We provided several arguments, some in the context of gravitoelectromagnetism (GEM), that this energy flux is, indeed, gravitational in nature. Moreover, this definition is simple, exact, and operational in nature---it can be measured by RQF observers using accelerometers and gyroscopes.

Finally, to demonstrate the importance of this new gravitational energy flux, we considered an apparent paradox that arises in a simple electromagnetism problem in special relativity. The paradox is that the increase in electromagnetic energy inside a rigid, accelerating box cannot be accounted for by the Poynting flux alone. We need to add another flux---the gravitational energy flux, $\alpha \cdot \mathcal{P}$, to get the correct answer. The latter flux cannot be properly understood in the context of special relativity; it involves geometrical effects at the boundary of the system, namely, a coupling between the intrinsic ($\alpha$) and extrinsic ($\mathcal{P}$) geometry of the boundary, which is properly in the domain of general relativity.

In the future, we plan to use the RQF approach to analyze the problem of motion in more depth.  We expect it will prove invaluable for this purpose because the quasilocal nature of an RQF allows one to treat the problem of motion without any direct knowledge or assumptions about the fields in the interior volume of the system. In particular, for problems like gravitational self-force, we expect that RQFs will allow us to circumvent the singularity issues one normally encounters with a point particle assumption.

\section*{Acknowledgments}

This work was supported in part by the Natural Sciences and Engineering Research Council of Canada.

\end{document}